\documentclass[12pt,preprint]{aastex}
\usepackage{amsmath}

\begin{document}

\title{Hadronic origin of prompt high-energy emission of gamma-ray bursts revisited: in the case of a limited maximum proton energy}

\author{Kai Wang$^{1,2,4}$, Ruo-Yu Liu$^{2,5}$,  Zi-Gao Dai$^1$ and Katsuaki Asano$^3$}
\affil{$^{1}$School of Astronomy and Space Science, Nanjing University, Nanjing 210093, China;  dzg@nju.edu.cn \\
$^{2}$Max-Planck-Institut f\"ur Kernphysik, Saupfercheckweg 1, 69117 Heidelberg, Germany; ruoyu@mpi-hd.mpg.de\\
$^{3}$Institute for Cosmic Ray Research, The University of Tokyo, 5-1-5 Kashiwanoha, Kashiwa, Chiba 277-8582, Japan; asanok@icrr.u-tokyo.ac.jp\\
$^{4}$Department of Astronomy and Kavli Institute for Astronomy and Astrophysics, Peking University, Beijing 100871, China; kaiwang@pku.edu.cn\\
$^{5}$Deutsches Elektronen-Synchrotron (DESY), Platanenallee 6, D-15738 Zeuthen, Germany; ruoyu.liu@desy.de
}

\begin{abstract}
The high-energy ($>100\,$MeV) emission observed by \emph{Fermi}-LAT during the prompt phase of some luminous gamma-ray bursts (GRBs) could arise from the cascade induced by interactions between accelerated protons and the radiation field of GRBs. The photomeson process, which is usually suggested to operate in such a hadronic explanation, requires a rather high proton energy ($ >10^{17}$\,eV) for an efficient interaction. However, whether GRBs can accelerate protons to such a high energy is far from guaranteed, although they have been suggested as the candidate source for ultrahigh-energy cosmic rays. In this work, we revisit the hadronic model for the prompt high-energy emission of GRBs with a smaller maximum proton energy than the usually adopted value estimated from the Bohm condition. In this case, the Bethe-Heitler pair production process becomes comparably important or even dominates over the photomeson process. We show that with a relatively low maximum proton energy with a Lorentz factor of $10^5$ in the comoving frame, the cascade emission can still reproduce various types of high-energy spectrum of GRBs. For most GRBs without high-energy emission detected, the maximum proton energy could be even lower and relax the constraints on the parameters of GRB jet resulting from the fact of non-detection of GRB neutrinos by IceCube.
\end{abstract}


\section{Introduction}
High-energy gamma-rays have been discovered in a few tens of gamma-ray bursts (GRBs), spanning from the prompt emission phase to the afterglow phase, by the Large Area Telescope (LAT) on board the \emph{Fermi} satellite \citep{ackermann13a}.
The long-lasting LAT emission in the afterglow phase, detected up to thousands of seconds or even one day after the burst, usually shows a power-law decay with time, and may arise from the external forward shock via synchrotron radiation and inverse Compton (IC) scattering of accelerated electrons in forward shocks \citep{kumar09, Kumar10, Wang13, liu13, fukushima17}. On the other hand,
in many of these GRBs, the high-energy emission during the prompt phase presents a rapid variability and a temporal correlation with the keV/MeV emission, implying an internal dissipation origin \citep{maxham11,Tang17}, rather than an external shock origin. Generally, the GRB spectra should contain three elemental components, i.e., Band component, blackbody component, and an extra power-law component extending to high energies \citep{zhang11b,guiriec15}. For the high-energy emissions, based on their spectra, one has three types, say,
(i) 090926A-type: the high-energy spectrum is hard (photon index $\sim -1.6$) and shows an evident high-energy cutoff \citep{ackermann11}, also forming an extra component;
(ii) 090902B-type: the high-energy spectrum is quite flat (photon index $\sim -2$) without an obvious high-energy cutoff in the Fermi/LAT energy region (the high-energy cutoff maybe exist at a higher energy beyond the upper limit of Fermi/LAT) \citep{abdo09a}, manifesting itself as an extra component relative to the keV/MeV one;
(iii) 080916C-type: the high-energy spectrum seems to be consistent with the extrapolation of the empirical Band function from the keV/MeV emission \citep{abdo09b,band93}.
There are also some common features shared by these GRBs, such as the high isotropic energy and large bulk Lorentz factor inferred from observations.

The radiation mechanism of the prompt high-energy emission is still unclear. Both leptonic and hadronic origins have been suggested. The leptonic origin usually relates to the external shock, such as synchrotron radiation of electrons accelerated in an external shock, or prompt keV/MeV photons up-scattered by the accelerated electrons in the external shock \citep{kumar09, ackermann11, beloborodov14}.
\citet{asano10} considered a proton-dominated GRB jet in which protons are accelerated up to $10^{20}$eV, initiating an electromagnetic cascade via the photomeson process in the photon field of the jet. These authors found that the intensity of secondary $e^{\pm}$ pairs produced in the cascade can significantly exceeds that of the  primary electrons at high energies. And their synchrotron radiation and the inverse Compton (IC) scattering off keV/MeV photons in the jet can reproduce the observed high-energy spectrum. Note that the photomeson process has a high threshold energy of $\sim 0.34$\,GeV for photon (in the rest frame of a proton), which translates to a proton energy of $E_p>10^{17} (\Gamma/300)^2(\varepsilon_{\gamma,b}/300\, {\rm keV})^{-1}\,$eV in the observer's frame, where $\Gamma$ is the bulk Lorentz factor of the jet, $\varepsilon_{\gamma,b}$ is the break energy in the spectrum of the prompt keV/MeV emission. Although the maximum proton energy can easily reach such a high energy when considering the acceleration by a relativistic shock in the Bohm condition especially for those very energetic GRBs, however, such a condition may not be achieved in real shocks. What's more, the energy dissipation mechanism during the prompt emission phase is actually unclear. Besides the shock acceleration, mechanisms such as neutron decay and magnetic reconnection have been also suggested to account for the conversion from the kinetic energy of the GRB jet to non-thermal energies of emitting particles \cite[e.g.,][]{Rees05, Zhang11}. Thus, the maximum achievable proton energy in GRB's jet is far from certainty.  In the case that the maximum proton energy falls below the threshold energy of the photomeson process to interact with photons of energy of the spectral peak, the photomeson process may not be efficient to produce the observed high-energy flux in a GRB. On the other hand, the Bethe-Heitler $e^\pm$ pair production process of protons ($p\gamma\to pe^+e^-$) has a lower threshold energy about 1\,MeV and its energy loss rate peaks at $\sim 10$\,MeV. Thus, protons may still induce an electromagnetic (EM) cascade via this process, producing high-energy emission during the prompt emission phase.

In this paper, we revisit the hadronic model for the prompt high-energy emission of GRBs, under the consideration of a relatively limited acceleration ability of protons. Different from the previous numerical studies, we are dedicated to an analytical method which may help us to better understand the physical processes underlying the observed radiation. This paper is organized as follows: in Section 2, the model and method adopted are described, and the results in the benchmark case are presented. We show that the spectrum of the prompt high-energy emission of GRB~090926A, GRB~090902B and GRB~080916C, which are regarded as representatives of three different types of the high-energy spectrum, can be well fitted in the considered case in Section 3. In Section 4, we discuss the possible leptonic contribution and neutrino emission under the considered model and we summarize this work in Section 5.

\section{Proton-induced cascade in the GRB jet}
Let us consider an isotropically expanding shell with the Lorentz factor $\Gamma$ at radius $R$ from the central engine. For simplicity, all the physical quantities inside the shell is assumed to be homogeneous. The spectrum of keV-MeV photons in the prompt emission phase can be usually depicted by the so-called Band function \citep{band93},
\begin{equation}
n(\varepsilon_{\gamma}) = \left\{ \begin{array}{ll}
  A{\left( {\frac{\varepsilon_{\gamma}}{{100{\text{keV}}}}} \right)^\alpha }\exp \left( { - \frac{\varepsilon_{\gamma}}{{{\varepsilon_{\gamma,0}}}}} \right), & \varepsilon_{\gamma} < (\alpha  - \beta ){\varepsilon_{\gamma,0}}, \hfill \\
  A{\left( {\frac{{(\alpha  - \beta ){\varepsilon_{\gamma,0}}}}{{100{\text{keV}}}}} \right)^{\alpha  - \beta }}\exp \left( {\beta  - \alpha } \right){\left( {\frac{\varepsilon_{\gamma}}{{100keV}}} \right)^\beta }, & \varepsilon_{\gamma} > (\alpha  - \beta ){\varepsilon_{\gamma,0}}, \hfill \\
\end{array}  \right.
\end{equation}
where $\alpha$ and $\beta$ are the low-energy and high-energy photon indexes, respectively, separated by the break energy $\varepsilon_{\gamma,b}\equiv (\alpha-\beta)\varepsilon_{\gamma,0}$.  The normalized coefficient is $A = {\Gamma^2U_\gamma }/\left[ {\int_{{\varepsilon _{\gamma ,\min }}}^{{\varepsilon _{\gamma ,\max }}} {n({\varepsilon _\gamma }){\varepsilon _\gamma }d{\varepsilon _\gamma }} } \right]$, where $U_\gamma =L_\gamma /(4 \pi R^2 \Gamma^2 c)$ is the photon energy density in the comoving frame,  and $L_\gamma$ is the luminosity integrated from $\varepsilon _{\gamma ,\min }$ to $\varepsilon _{\gamma ,\max}$, which are fixed to be $1\,\rm keV$ and $10\,\rm MeV$ respectively.

In the outflow, the primary protons are assumed to be accelerated to a power-law distribution in energy space, say, ${{n'}_p}({{\gamma '}_p}) \propto {\gamma'}_p ^{-s}$ for $\gamma'_{p,\min} \leqslant \gamma'_{p} \leqslant \gamma'_{p,\max}$, where $\gamma'_{p,\min}$ is taken to be just slightly larger than unity in the comoving frame and $\gamma'_{p,\max}$ is treated as a parameter because of its uncertainty. The energy density of the accelerated protons in the comoving frame of the outflow is linked with the photon density by $U_p=\epsilon_p U_\gamma$ \footnote{note that the definition here is not identical to the normal definition of the energy equipartition coefficient, which is usually the fraction of the dissipated kinetic energy that goes in to nonthermal protons.}. As relativistic protons are injected into the dense keV/MeV photon field of the outflow, the photomeson process and the BH process would operate and produce the secondary gamma-ray photons, $e^\pm$ pairs, and neutrinos. For both processes, we adopt a semi-analytical treatment for the emissivity and spectrum of the secondaries, following \citet{kelner08}.

\begin{figure}
\epsscale{.96} \plotone{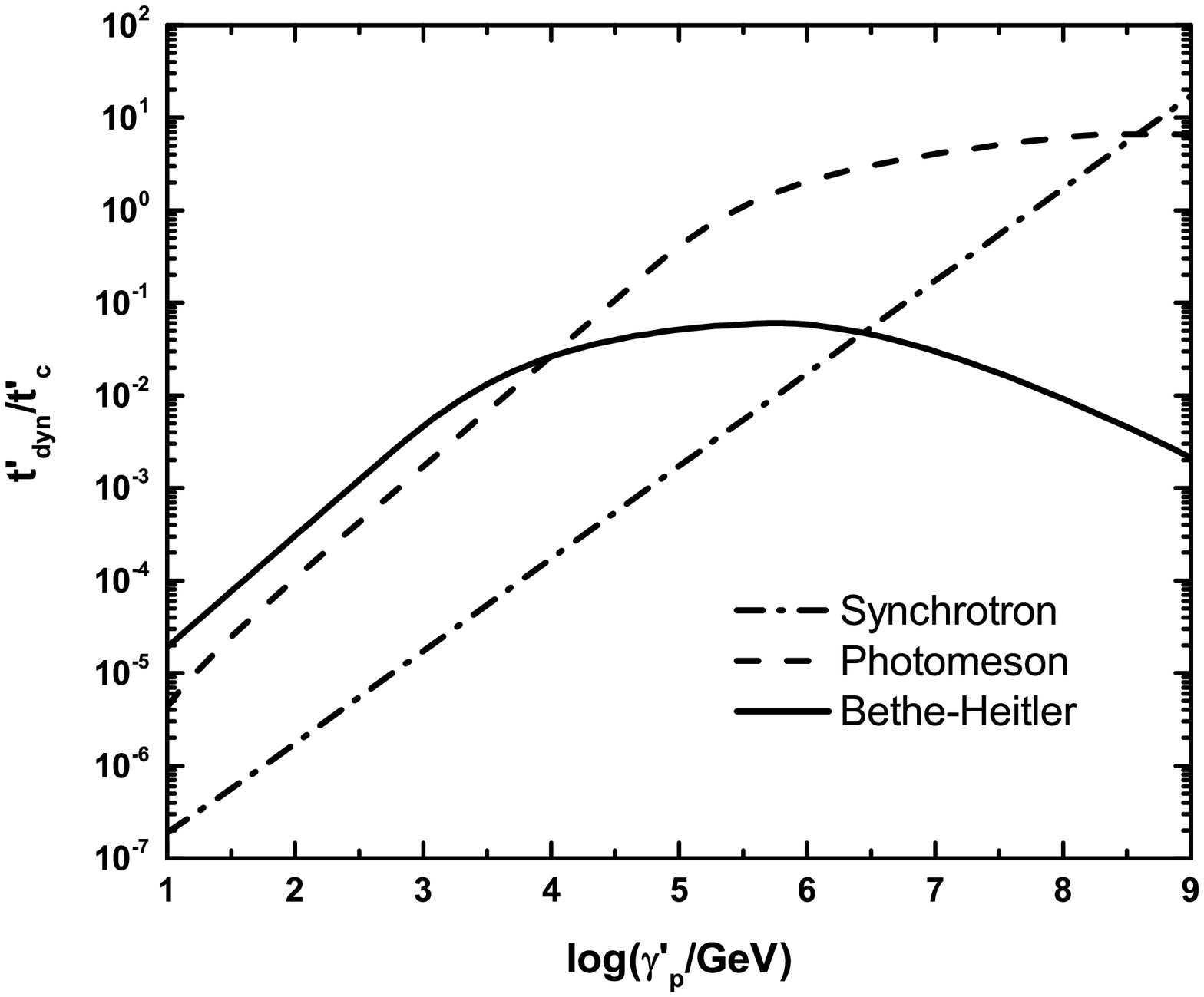} \caption{The ratios between the dynamical timescale and the cooling timescales of different processes for a proton as a function of the proton's Lorentz factor. All the quantities are measured in the comoving frame and calculated with the benchmark parameters for GRB jet.\label{fig:timescale}}
\end{figure}

In our calculations, the free parameters are the luminosity $L_\gamma$ in 1\,keV$-$10\,MeV, bulk Lorentz factor of the outflow $\Gamma$, the dissipation radius $R$, the energy equipartition coefficients for protons $\epsilon_p$ and for magnetic field $\epsilon_B$, the spectral properties of the keV/MeV emission, say, spectral index $\alpha$, $\beta$ and the break energy $\varepsilon_{\gamma,b}$, the redshift of the GRB $z$ and the maximum proton energy (Lorentz factor) in the comoving frame $\gamma_{p,\max}'$ (or the Bohm factor) and the spectral index $s$. Although GRBs detected by LAT during the prompt emission phase usually show a large bulk Lorentz factor, we consider a moderate one, say, $\Gamma=300$, as the benchmark value since it is a more commonly accepted value for most GRBs. The benchmark values of other parameters can be found in Table.~1. Fig.~1 presents the ratio between the dynamical timescale of one pulse in the comoving frame of the outflow $t'_{\rm dyn}\simeq R/\Gamma c$ and the energy loss timescales $t'_{\rm c}$ of relevant processes in the comoving frame as a function of proton energy under the benchmark parameters. These ratios can be regarded as the cooling efficiency of these processes, i.e., $t_{\rm dyn}'/t_{p\gamma}$ for photomeson cooling efficiency, $t_{\rm dyn}'/t_{\rm BH}$ for BH cooling efficiency and $t_{\rm dyn}'/t_{\rm syn}$ for synchrotron cooling efficiency. Such results shown in Fig.~1 are consistent with the estimations in the earlier studies \citep[e.g.][]{crumley13,kumar15}, i.e., the BH cooling dominates over the photomeson cooling at lower $\gamma_{p}'$ while at larger $\gamma_{p}'$ the latter one is dominant. Note that the energy loss timescale for the photomeson and the BH process depends on the cross section of the interaction (denoted by $\sigma$) and the fraction of energy lost by proton in one interaction (i.e., the inelasticity $\kappa$), given a fixed number density of the target photon field. The value of the product of these two quantities $\kappa\sigma$ reaches the peak when the photon energy is $0.3\,$GeV \citep{Mucke00} and $10\,$MeV \citep{chodorowski92} in the rest frame of the proton for both the photomeson process and the BH process respectively. Thus, for each proton energy, there is a corresponding typical photon energy $\varepsilon_{\rm t}$, at which the interaction or the energy loss will be the most efficient than that at other energies. We obtain $\varepsilon_{\rm t}\simeq 1\,\rm MeV (\Gamma/300)(\gamma_p'/10^5)^{-1}(1+z)^{-1}$ for the photomeson process and $\varepsilon_{\rm t}\simeq 30\,\rm keV (\Gamma/300)(\gamma_p'/10^5)^{-1}(1+z)^{-1}$ for the BH process. Therefore, for a given spectral break at $\varepsilon_{\gamma,b}=300\,$keV with $z=1$ in the benchmark case, we can expect a break in the curve of the cooling timescale as a function of $\gamma_p'$ for the photomeson process around $\gamma_p'=1.5\times 10^{5}$, and that for the BH process around $\gamma_p'=5\times 10^{3}$, as can be seen in Fig.~\ref{fig:timescale}. The second break for the BH process around $\gamma_p'=10^{6}$ is due to the low energy cutoff of the Band spectrum at $\varepsilon_{\min}=1\,\rm keV$, because the corresponding $\varepsilon_{\rm t}$ for higher-energy protons is even lower than the minimum energy of the target photon field.

In addition to the photomeson process and the BH process, synchrotron radiation can also make protons lose their energies.
The synchrotron cooling timescale for relativistic proton is $t'_{syn} = \frac{9}{4\gamma_p'}\frac{m_p^3c^5}{e^4B^2}$, which is less important than either the photomeson or the BH processes for $\gamma'_{p} < 3\times 10^{8}$. We ignore the inverse Compton cooling of proton in this work since it would be dominated by the photomeson cooling at any energies \citep{asano05}. Usually,in terms of $\gamma_{p,\max}'$, it can be determined by equating the acceleration timescale ${{t}_{acc}'} = \mathcal{A}{t_L}\beta_s^{-2} \simeq \mathcal{A}{{\gamma '}_p}{m_p}c/eB$ to $\min \{ {{t}_{dyn}'},{{t}_{c}'}\} $, where $\beta_s$ is the speed of the internal shock in unit of the speed of light which is close to unity in the scenario of GRB internal shock, ${t'_L}$ is the Larmor gyration timescale, ${{t'}_{c}}$ is the energy loss time due to the photomeson, BH and synchrotron, and $\mathcal{A}(>1)$ is defined as the Bohm factor which measures the deviation from the acceleration in the Bohm limit. In this work, we are mainly concerned with $\gamma_{p,max}'\leq 10^6$ ($\mathcal{A} \gg 1$), where the energy loss efficiency are smaller than unity. Thus, as an approximation, we consider a constant injection rate of secondary particles from the photomeson process and the BH process during one pulse.


Once being produced, the secondary photons and $e^\pm$ pairs of sufficiently high energies inevitably initiate an EM cascade in the dense keV/MeV photon fields via $\gamma\gamma$ annihilation and IC scattering. High-energy photons can also be produced via synchrotron radiation of $e^\pm$ pairs in the strong magnetic field, given $B\simeq 10^5 \epsilon_{B}^{1/2}(L_{\gamma}/10^{53}\,\rm erg/s)^{1/2}(R/10^{14}\,\rm cm)^{-1}(\Gamma/300)^{-1}\,$G. For simplicity, the synchrotron radiation of intermediate particles such as charged pions and muons are treated as the suppression factors, respectively, $1-\exp( - {{t'}_{\pi,syn}(E_\pi')}/{{\tau '}_\pi }(E_\pi'))$ and $1-\exp( - {{t'}_{\mu,syn}(E_\mu')}/{{\tau '}_\mu(E_\mu') })$ which rely on energy of the parent proton, say, $E_\pi =0.2 E_p$ and $E_\mu =0.15 E_p$ where ${\tau '}_\pi =2.6 \times 10^{-8}\gamma'_\pi\,\rm s$ and ${\tau '}_\mu =2.2 \times 10^{-6}\gamma'_\mu\,\rm s$ are the lifetimes of pions and muons. On the other hand, in the considered energy range ($\gamma_{p,\max}'\leq 10^6$), the synchrotron radiation of these intermediate charged particles is very weak comparing to contribution from secondary electrons and hence their contribution to the overall flux is neglected. The differential density of the cascaded electron (including both $e^-$ and $e^+$) $n_e(\gamma)$ is governed by the energy transport equation
\begin{equation}
\frac{\partial n_e'}{\partial t}+\frac{\partial}{\partial \gamma_e'}\left(\dot{\gamma'_e}n_e'\right)=Q_e+\dot{n}_{e,\gamma\gamma}',
\end{equation}
where $Q_e$ is the injection of electrons from protons via the photomeson process or the BH process while $\dot{n}_{e,\gamma\gamma}'$ represents the injection of electrons from photons via the $\gamma\gamma$ annihilation. We ignore the escape term in the above equation since electrons can hardly escape such a highly turbulent outflow of GRBs with a strong magnetic field and a very dense photon field. The cascade develops quite quickly and the distribution of electrons in energy space would reach the quasi-steady state since the timescales of relevant processes are typically much shorter than the dynamical timescale $ t_{\rm dyn}'$ in the GRBs' environment. Thus, we can derive the electron spectrum from the energy transport equation in the quasi-steady state ($\partial n_e'/\partial t=0$) by
\begin{equation}
{{n'}_e}({{\gamma '}_e}) = -\frac{1}{{{\dot{\gamma}'}_e}}\int_{{{\gamma '}_e}}^\infty  d {{\tilde \gamma }_e'}[{{Q_e}({{\tilde \gamma}_e'}) + {\dot n'}_{e,\gamma \gamma } ({{\tilde \gamma }_e'})}],
\label{eq:e_spec}
\end{equation}
and avoid the time-consuming Monte Carlo simulations of the development of the cascade. Our detailed treatment can be found in the Appendix.

In Fig.~\ref{fig:espec}, we show the cascaded electron spectrum in the quasi-steady state with different $\gamma_{p,\max}'$. We separate the electrons originating from the BH process (solid curves) from those originating from the photomeson process (dashed curves). Basically, the spectrum from these two processes have a similar structure, with a power-law of $\sim \gamma_e'^{-3}$ at low energies, becoming harder at certain energy and then followed by a cutoff. The low energy part is mainly determined by the electrons produced in the $\gamma\gamma$ annihilation. At such an energy, the produced $e^\pm$ pair approximately shares the energy of the parent photon and the injected electron spectrum generally follows the spectrum of the high-energy photon provided by a high interaction rate in the outflow, which can be assumed to be of the form $E^{-\Gamma_\gamma}$. Based on Eq.~(\ref{eq:e_spec}), the power-law index of electrons in the quasi-steady state is then $\Gamma_e=\Gamma_\gamma+1$. On the other hand, high-energy photons should be produced via either synchrotron radiation or the inverse Compton radiation by those electrons (the decay of $\pi^0$ from photomeson process is very high and makes a negligible contribution of gamma-ray at the considered energy here), giving rise to the relation $\Gamma_\gamma=(\Gamma_e+1)/2$. Thus, we obtain $\Gamma_e=3$ and $\Gamma_\gamma=2$, which is consistent with our results shown in Fig.~\ref{fig:espec} and Fig.~\ref{fig:gmax}. We note that in some scenarios of the EM cascade (such as the one in the hadronic model for GeV photons of blazars, or the one developed in the intergalactic space), a flat electron spectrum ($\Gamma_e=2$) is expected to appear at the low energy, which is due to the cooling of higher-energy electrons without injection. In the scenario of a GRB jet, however, the injection of electrons from $\gamma\gamma$ annihilations can extend to quite low energies since the energy of target photons is high, and hence such a segment of the spectrum will not show up. At higher energies, the injection of electrons from the channel of $\gamma\gamma$ annihilation becomes less important due to insufficient high energy photons, while the photomeson process and the BH process becomes more efficient. Thus, the electrons injected directly from the interactions of protons cause the hardening of the quasi-steady-state electron spectrum at high energies. The cutoff in the electron spectrum is due to the cutoff in the proton spectrum. Because the energy of secondary electron produced in the photomeson process is higher than that in the BH process, the cutoff energy of the electron spectrum from the photomeson process is higher than that from the BH process.

The spectrum of the cascade emission with different $\gamma_{p,\max}'$ is exhibited in Fig.~\ref{fig:gmax}. In the case of $\gamma_{p,\max}'=10^5$ or the benchmark case, the contribution of the BH process is higher than that of the photomeson process below $1\,$GeV. Note that in the benchmark case, the typical energy of electrons producing 1\,GeV photon via synchrotron radiation is $\gamma_e'\simeq 10^4$, while the Lorentz factor of the electrons produced directly from the BH process is typically $\sim 10^{-3}(m_p/m_e)\gamma_p'\simeq \gamma_p'$. So this energy is consistent with the proton energy below which the BH process is more efficient than the photomeson process of cooling protons (as shown in Fig.~\ref{fig:timescale}). The spectral breaks around 100\,MeV is due to $\gamma\gamma$ annihilation inside the jet. Note that the photomeson efficiency ($\sim t'_{\rm dyn}/t'_{p\gamma}$) is about 8 times higher than the BH efficiency ($\sim t'_{\rm dyn}/t'_{\rm BH}$) at $\gamma_p=10^5$, as can be seen in Fig.~\ref{fig:timescale}. Considering a proton spectrum with an index $s=2$ and $\gamma_{p,\max}'=10^5$, the overall photomeson efficiency (integral over the proton spectrum) is still about 4 times higher than the BH efficiency. After subtracting the energy taken away by the generated neutrinos, the energy of protons lost to EM components from the photomeson process is still about twice that from the BH process. Thus, one may expect the photomeson process have a larger contribution here. However, we should keep in mind that the photons we observe are after absorption while protons with the same energy generally produce higher energy electrons via the photomeson process than via the BH process. Before the absorption of high energy photons due to $\gamma\gamma$ annihilation, the spectrum of the cascade emission from the photomeson process peaks around TeV while that from the BH process peaks at several tens of MeV. The peak flux of the former one is indeed about twice that of the latter one. However, the contribution of the BH process is more important than that of the photomeson process around 100\,MeV. After the absorption of the TeV gamma-rays, energy are reprocessed into lower energies photons (below MeV) via synchrotron radiation of generated secondary electrons. This explains why the BH process dominates around 100\,MeV although its overall efficiency is less than that of the photonmeson process.

In Fig.~\ref{fig:gmax}, we can also see the flux of the cascade emission is higher with a larger $\gamma_{p,\max}'$. This is because the interaction efficiency for a proton increases with the proton's energy, especially for the photomeson process. So a large $\gamma_{p,\max}'$ enhances the overall interaction efficiency and consequently the flux of the cascade emission. This is the reason that the cascade flux is reduced by a lot in the case of $\gamma_{p,\max}'=10^4$. What's more, in the case of $\gamma_{p,\max}'=10^4$, the maximum Lorentz factor of an electron is $\sim 10^{-3}m_p/m_e\sim \gamma_{p,\max}'=10^4$. This leads to the softening of the spectrum at a few MeV. Thus, we do not expect the hadronic model can work well in explaining GeV emissions with such a low maximum energy for protons and hence do not explore the case of $\gamma_{p,\max}'=10^4$ in the fitting of the spectrum of LAT GRBs. But this case would be a hint for the non-detection of neutrinos as we will discuss later. In the case of ${\gamma'}_{p,\max} \gtrsim 10^6$, the BH process contributes to the total emission by a few tens of percent, which is consistent with the claim by \cite{asano09a}. In the figure, each curve for the BH process and photomeson process is the sum of the synchrotron radiation and the IC radiation by the cascaded $e^\pm$ pairs. We note that whether the photomeson process has more contribution than the BH process does not only depend on $\gamma_{p,\max}'$. The spectrum shape of the target photon field in the comoving frame, in terms of the break energy $\varepsilon_{\gamma,b}'(\simeq \varepsilon_{\gamma,b}/\Gamma)$ and low-energy spectral index $\alpha$, also has important influence. In the benchmark case, the break energy is $\varepsilon '_{\gamma,b}=2$\,keV in the comoving frame. If the break energy is higher, the turnover in both photomeson efficiency around $\gamma_p'=10^5$ as shown in Fig.~\ref{fig:timescale} will shift to lower energy and hence the photomeson process will dominate even in the case of $\gamma_{p,\max}'=10^5$ \citep{Gao12}. On the other hand, if the low-energy spectrum of the Band component is softer (i.e. a smaller $\alpha$), the photomeson efficiency would not decrease so fast with energy below the turnover and may still have more contribution than the BH process.

We also investigate the influence of the bulk Lorentz factor $\Gamma$ and the dissipation radius $R$. The results are shown in Fig.\,\ref{fig:LF} and Fig.\,\ref{fig:radius}. Due to the energy density of target photon field has the dependence $U_\gamma  \propto \Gamma^{-2}R^{-2}$, a larger $\Gamma$ or $R$ results in a lower interaction efficiency, so the flux of cascade emission decreases as $\Gamma$ and $R$ increase, and the cutoff energy due to $\gamma\gamma$ annihilation increases for a larger $\Gamma$ and $R$. 

In Fig.~\ref{fig:epB}, the effect of the equipartition coefficient for magnetic field $\epsilon_B$ {which is defined as the ratio of magnetic field energy density to the photon energy density in the comoving frame (i.e. $\epsilon_B\equiv U_B/U_\gamma$)}, on the spectrum of the cascade emission is shown. {The magnetic field in the comoving frame is then $B=\sqrt{2\epsilon_BL_\gamma/\Gamma^2R^2c}$}. Basically, a higher $\epsilon_B$ leads to a more important synchrotron radiation, while a smaller $\epsilon_B$ leads to a more important IC radiation. Since electrons usually emit higher energy photons via the IC radiation than via the synchrotron radiation, the flux of the cascade emission at high energies becomes higher for a smaller $\epsilon_B$. On the other hand, more high energy photons in turn produce more $e^\pm$ pairs at high energies, making a harder electron spectrum than that in the case of a large $\epsilon_B$, as is shown in Fig.~\ref{fig:espec_epB}. This explains why the spectrum is also harder in the case of $\epsilon_B=0.01$ than that in the $\epsilon_B=1$ case at keV energies, where the emission are dominated by synchrotron radiation of these (relatively) low-energy electrons in both cases. One can see that the gamma-ray flux around 100\,MeV is not very sensitive to the value of $\epsilon_B$. This is because that the injection rate of the secondary electrons which emit 100\,MeV photons is dominated by the $\gamma\gamma$ annihilation process (i.e., $\dot{n}_{e,\gamma\gamma}$ in the r.h.s. of Eq.~\ref{eq:e_spec}), and mainly relies on the photon field and does not change with the magnetic field. On the other hand, the jet is already an electron calorimeter even if only considering IC radiation under the benchmark parameters. Therefore, we expect the total photon production rate at certain energy is roughly equal to the energy injection rate of the electrons which emit these photons (i.e., $L_{\gamma}\sim L_e$). A large $\epsilon_B$  only changes the way the electrons radiate their energy away but not increase the photon production rate. Although the energy of the gamma-ray emitting electrons is different in the IC-dominated case with a small $\epsilon_B$ from in a synchrotron-dominated case with a large $\epsilon_B$, the resultant gamma-ray flux is at the the same level, since that the energy injection rate of electrons in the relevant energy range is roughly constant with respect to electron energy (i.e.,  $E^2\dot{n}_{e,\gamma\gamma}\propto E^{0}$).

\begin{center}
\begin{table}
\caption{The benchmark parameters of GRB outflow. \label{tab:benchmark}}
\begin{tabular}{ccccc}\hline
Parameter & Values  \\\hline
$z$ 	& 1  \\
$\Gamma$  	& 300  \\
$s$  	& 2  \\
$\alpha$  	& -1  \\
$\beta$  	& -2.2 \\
$\varepsilon_{\gamma,p}$  	& $300\,\rm keV$  \\
$L_\gamma$(1\,keV--10\,MeV)  & $10^{53}\,\rm erg/s$  \\
$\epsilon_p$  & 10  \\
$\epsilon_B$  & 1 \\
$R$  & $10^{14}\,\rm cm$ \\
$\gamma'_{p,\max}$  & $10^5$ \\
\hline
\end{tabular}
\end{table}
\end{center}

\section{Application to LAT GRBs}
In this section, we apply our calculations to explain the high-energy spectrum of GRB~090926A, GRB~090902B and GRB~080916C, which represent three different types of spectral feature. The origin of prompt keV/MeV emission is beyond the scope of this work and we just regard it a pre-existed target photon field for the BH process and the photomeson process. In our calculations below, we employ a flat spectrum for the injected proton spectrum ($s=2$). Since the efficiencies for both photomeson and BH process increase with proton energy in the considered energy range, the gamma-ray flux (also the neutrino flux) will be lower if we employ a softer proton spectrum, and vice versa. However, the shape of the spectrum of the cascade emission should not change significantly, as the spectrum of secondary $e^\pm$ does not depend significantly on the spectrum at injection as long as the cascade has fully developed. Thus, what mainly matters is the baryonic loading factor.

\subsection{GRB~090926A}\label{090926a}
GRB~090926A, a luminous long GRB at redshift $z=2.1$, was detected at high-energy (above $100\,$MeV) by \emph{Fermi}-LAT during the prompt phase with both a spatial and temporal correlation with the \emph{Fermi}-GBM data \citep{ackermann11}. The spectrum in GBM band is selected by a time-integrated spectrum within an interval between $T_0+3.3\,$s and $T_0+21.6\,$s, based on the Band function with the best fitting parameters of $\alpha= -0.6$, $\beta = -2.6$ and $\varepsilon_{\gamma,p} =256\,\rm keV$. The high-energy spectrum manifest itself as an extra component in addition to the conventional Band component in both the time-integrated and the time-resolved spectra with an evident spectra cutoff around GeV. The peak of the GeV lightcurve and that of the Band component are in coincidence, implying a strong correlation of the origin of these photons.

The spectral fit to GBM and LAT data by the Fermi Collaboration is present by the green dashed lines in Fig.~\ref{fig:090926A}. We take the keV/MeV component as the target photon and calculate the EM cascade emission induced by the  photomeson production and BH process to see whether it can match the observed high-energy extra component. Due to the hard spectrum (with a photon index $-1.72$) of the extra component, the IC radiation is supposed to play the leading role around GeV. As a result, a small $\epsilon_B$ is expected. We adopt two maximum Lorentz factors of accelerated protons, i.e., $\gamma'_{p,\max}=10^5$ and $\gamma'_{p,\max}=10^6$, and the two corresponding Bohm factors are $\mathcal{A} \simeq 830$ and $\mathcal{A} \simeq 67$ for two groups of parameters adopted in Fig.~\ref{fig:090926A}, respectively. In the former case, high-energy emission is dominated by the BH processes while in the latter one the photomeson process dominates. Both cases can give a good fit to the data, and requires high baryonic loading factor ($\epsilon_p > 10$), consistent with the result in previous studies on the hadronic origin of the extra component of GRBs \citep{asano09a,asano09b}. Note that a larger proton energy budget requires to be invoked for $\gamma'_{p,\max}=10^5$ owing to a lower overall interaction efficiency for proton.

Due to the measured cutoff in the high-energy spectrum, the Lorentz factor of the outflow is estimated to be $\Gamma \simeq 720 \pm 76$ in the framework of internal shock model \citep{ackermann11}. In addition, the radius is estimated to be $10^{14}-10^{15}\,\rm cm$. In our fitting, with the radius $R=10^{14}\,\rm cm$ being fixed, a Lorentz factor of $\Gamma=720$ and $\Gamma=800$ is adopted to produce spectrum cutoff for $\gamma_{p,\max}'=10^5$ and $\gamma_{p,\max}'=10^6$ respectively, which is consistent with the estimation in \cite{ackermann11}.

\subsection{GRB~090902B}
GRB~090902B is another intense burst detected at a redshift of $z=1.822$. The joint spectral fit of GBM and LAT data between $T_0+4.6\,$s and $T_0+9.6\,$s by \cite{abdo09a} shows the spectrum can be described by a Band function of $\alpha = 0.07$, $\beta = -3.9$, and $\varepsilon_{\gamma,p} = 908\,\rm keV$ on top of a power-law  with a photon index $-1.94$, with excesses at both high energy ($>100\,\rm MeV$) and low energy (below $50\,\rm keV$). The low-energy excess is not a unique feature to GRB~090902B, and there are some other GRBs presenting the same feature, e.g., GRB~080319B and
GRB~110731A \citep{asano10,ackermann13b}. For GRB~090902B, the low-energy excess seems to be consistent with the extrapolation from the high-energy excess, and can be reproduced by the cascade emissions induced by the photomeson process along with the high energy one \citep{asano10}. In addition to the target photons from Band spectrum, the contributions to target photons from the low-energy excess is taken into account as well by implementing the iteration.


Since the extra high-energy spectrum extends up to tens of GeV without a clear cutoff, a lower limit of $\Gamma=1000$ is inferred for the bulk Lorentz factor of the GRB outflow \citep{abdo09a}. Such a large $\Gamma$ would lead to a low efficiency for the interaction processes hadronic process, and requires a high baryonic loading factor to explain the extra component. Similar to the fitting to GRB~090926A, we also adopt  $\gamma'_{p,\max}=10^5$ and $\gamma'_{p,\max}=10^6$ for GRB~090902B, and the two corresponding Bohm factors are $\mathcal{A} \simeq 3500$ and $\mathcal{A} \simeq 760$ for two groups of parameters adopted in Fig.~\ref{fig:090902B}, respectively. As is shown in Figs.~\ref{fig:090902B}, the BH process dominates in the former case while the photomeson process dominates in the latter case. For the best fitting parameters, both cases require $\epsilon_p>100$, resulting in an isotropic-equivalent proton luminosity beyond $10^{55}\,\rm erg/s$. Such a large proton luminosity is normally a challenge to the hadronic model. However, since GRBs with extra components are usually very luminous ones, a high proton luminosity may be possible. On the other hand, the beaming correction may alleviate the energetic problem by a factor of $\theta^2/2$ with $\theta$ being the jet opening angle.

Fixing the dissipation radius at $R=10^{14}\,$cm, we reproduce the high-energy excess by adopting $\Gamma=1100$ and $1500$ for $\gamma_{p,\max}'=10^5$ and $\gamma_{p,\max}'=10^6$ respectively. In the case of $\gamma_{p,\max}'=10^5$, the BH process dominates in this case, and due to a small maximum proton Lorentz factor, the Lorentz factors of secondary $e^\pm$ are generally $\lesssim 10^5$. Thus, their synchrotron radiation can not produce GeV photons and a small $\epsilon_B=0.06$ is invoked to make sure efficient IC emission at GeV. The superposition of synchrotron and IC emission is used to explain the flat extra high energy component. Although they are somewhat fine-tuning, all the parameters are in their reasonable ranges. For the case of $\gamma'_{p,\max}=10^6$ (photomeson dominated case), the Lorentz factors of produced $e^\pm$ could reach $10^8$. Thus, synchrotron radiation of these electrons can solely account for the extra component given an appropriate value of $\epsilon_B$. The low-energy excess below 50\,keV is roughly fitted in both two cases, but with a slight difference in the slope. As we analyzed in the previous section and show in Fig.~\ref{fig:espec_epB}, this is because that the spectrum slope is related to $\epsilon_B$ at keV energies where the synchrotron radiation of cascaded electrons dominates. The value of $\epsilon_B$ in the case of $\gamma_{p,\max}'=10^5$ is smaller than that in the case of $\gamma_{p,\max}=10^6$, so the spectrum at keV energies in the latter case is softer than that in the former case.

The narrow Band component in this GRB may imply the photospheric emission \citep{zhang11b}. In this case, the emission radius may be smaller than what we used for fitting. However, when we adopt a smaller radius and larger $\Gamma$ (to avoid a too strong $\gamma\gamma$-absorption), the low-energy spectrum at $\sim 10$ keV becomes too hard to agree with the observed low-energy excess.

We note that, our fitting result in a large baryonic loading factor of $\epsilon_p>100$, while in \cite{asano10}, they adopted $\epsilon_p=3$ only. One reason for such a difference is that their result did not really reach the flux level of the observed one above $\sim \rm GeV$. Another reason is that they used a very high maximum proton energy estimated in the Bohm limit, which enhances the overall efficiency for the photomeson process.

\subsection{GRB~080916C}
GRB~080916C is the most powerful GRB ever recorded with the highest inferred isotropic energy $E_{iso}=8.8\times 10^{54}$\,erg in the energy range of 10\,keV$-$10\,GeV, located at a large redshift $z=4.35$ \citep{abdo09b}. The time-integrated spectrum between $T_0+3.58\,\rm s$ and $T_0+7.68\,\rm s$, i.e., the interval 'b' in \cite{abdo09b}, is fitted by a single Band function with $\alpha = -1.02$, $\beta = -2.21$ and $\varepsilon_{\gamma,p} = 1170\,\rm keV$ from keV to GeV. A correlation in temporal behavior between the GBM and LAT emission is found in this GRB as well. Although a single Band-function seems to explain the entire spectrum of this GRB from keV to GeV, the poor statistics at GeV can not rule out a different origin for the GeV emission from the keV/MeV emission.


Similar to the treatment for the two bursts above, we consider the both cases for $\gamma_{p,\max}'=10^5$ and $\gamma_{p,\max}=10^6$, and the two corresponding Bohm factors are $\mathcal{A} \simeq 730$ and $\mathcal{A} \simeq 73$ under two groups of parameters adopted in Fig.~\ref{fig:080916C}, respectively. Although the soft spectrum at high energy may be consistent with a synchrotron origin, the value of $\epsilon_B$ or magnetic field for this burst cannot be too large, otherwise the strong synchrotron radiation of secondary $e^\pm$ will lead to a low-energy excess $\lesssim 10\,$keV which is not observed. Thus, we expect that IC radiation of secondary electrons is the main contributor of the high-energy photons.  By treating the observed photons as the target radiation field, the $\gamma\gamma$ annihilation opacity for the highest-energy observed photon implies a minimum bulk Lorentz factor of $\Gamma_{\min} \sim 900$ and a dissipation radius of $R \gtrsim 10^{15}\,\rm cm$ for the interval considered \citep{abdo09b,zhang09}. In our calculations, $\Gamma= 1500$ and $R = 10^{15}\,\rm cm$ are employed for the adopted $\gamma_{p,\max}'$. Such a large bulk Lorentz factor and a large dissipation radius yield a low efficiency for the interactions of protons, so a large proton energy budget is needed as in the case of GRB~090902B \citep{wang09}. The results are shown in Fig.~\ref{fig:080916C}.

Given a large break energy $\varepsilon_{\gamma,p} = 1170\,\rm keV$ and a hard spectrum after the break ($\beta = -2.21$), the photomeson process have more contribution than the BH process around GeV even in the case of $\gamma'_{p,\max}=10^5$. For different maximum proton energies, the photomeson process contributes more around GeV than the BH process. The flux of the cascade emission peaks around GeV and the flux level is consistent with the extrapolation of the Band component from MeV.

\section{Discussions}
\subsection{Synchrotron-self Compton radiation of primary electrons}
In our calculation, we neglect the possible contribution by primary electrons to high-energy flux. In principle,  we can expect electrons are accelerated to relativistic energies in a GRB jet and scatter off photons from the Band component up to GeV--TeV range, although the Klein-Nishina effect can suppress the flux to certain extent. In an early study \citep{gupta07}, the authors compared the hadronic emission and the synchrotron-self Compton (SSC) emission by primary electrons in GRB jets within different sets of parameter regimes, and they found the SSC emission could outshine the hadronic emission under certain conditions. We here check the SSC contribution of the primarily accelerated electrons under the adopted parameters in the above fittings.

The synchrotron radiation of non-thermal electrons may contribute flux at keV/MeV range. Although the synchrotron origin of the Band component suffers from a lot of criticisms \citep[see][for reviews]{Zhang11a, zhang16}, we normalize the synchrotron emission to the observed Band component here to obtain an upper limit for the amount of the primarily accelerated electrons, regardless of  the excess at low energy of GRB~090902B and GRB~090926A. We then calculate the SSC radiation from the obtained electrons and the initiated cascade emission following the same treatment to the hadronic one. Due to the KN effect, SSC emission is severe suppressed at high energy, which causes the spectrum break in the SSC component shown in Fig.~\ref{fig:ssc}. Note that the cascade spectrum is more or less universal for a given background photon field, as long as the cascade is initiated at a sufficiently high energy and fully developed. Thus, whether the proton-initiated cascade emission (i.e., hadronic emission) or the SSC-initiated cascade emission dominates is determined by the separate injection rate. For GRB~090902B, given a large $\epsilon_B$ and a large $\epsilon_p$, we expect the SSC-initiated cascade emission is much weaker than the proton-initiated cascade emission, which can be also seen in the figure. For GRB~080916C, the $\epsilon_B$ is much smaller which leads to a higher SSC intensity but the $\epsilon_p$ is still large enough, so that the hadronic emission still dominates. On the contrary, we have a small $\epsilon_B$ and a (relatively) small $\epsilon_p$ for GRB~090926A. As a result, the SSC-initiated cascade exceeds the hadronic one. Nevertheless, this does not rule out the hadronic origin of the high-energy emission of GRB~090926A because the result is based on the assumption that the Band component is synchrotron emission. For GRB~090926A, the low-energy photon index is harder than the so-called ``line of death'' of the synchrotron radiation (i.e., $-2/3$, \citealt{preece98}), implying some other origins, e.g., photospheric radiations and Comptonization \citep{meszaros00,medvedev00, peer12}. Nevertheless, there may still be electrons being accelerated associating with protons to a power-law distribution in GRB~090926A. We also check emission of such a component of electrons. By using the same Bohm parameter $\mathcal{A}$ and power-law index as for proton, and normalizing the total electron luminosity by $L_e=L_\gamma$ ({or the energy equipartition coefficient for electrons is $\epsilon_e\equiv U_e/U_\gamma=1$ with $U_e$ the energy density of nonthermal electrons in the comoving frame}), we obtain an electron spectrum and their cascade emission, as shown with blue solid curve in the leftmost panel of Fig.~\ref{fig:ssc}. To make the electron contribution less important than the hadronic one for the adopted parameters in GRB~090926A, $\epsilon_e < 1$ (or $L_e/L_p<1/15$) is required.  {Such a result is fairly expected since electrons are more efficient emitters than protons, and it is consistent with what was found in the previous study \citep{gupta07}.}

\subsection{neutrino emission}
Since the photomeson process produces high-energy neutrinos via the decay of charged pions and muons, GRBs have been considered as one of the most promising high-energy neutrino source. However, no significant correlation between high-energy neutrino events and GRB events has been detected by the IceCube neutrino telescope. This result indicates a low neutrino emission rate of GRB jets, either due to a low photomeson efficiency or a low baryonic loading factor, and hence put stringent constraints on the parameters of GRB jets, e.g., the baryonic loading factor $\epsilon_p$, the bulk Lorentz factor $\Gamma$ and the dissipation radius $R$ \citep{aartsen15,aartsen16, aartsen17}. Note that the parameters adopted in our fittings to GRB~090926A, GRB~09092B, and GRB~080916C are in the allowed region. In our fittings, we adopt two maximum proton energies, i.e., $\gamma_{p,\max}'=10^6$ and $\gamma_{p,\max}'=10^5$. In the former case, the high-energy prompt emissions mainly arises from the photomeson process, and the peak flux of neutrinos is comparable to that of the photons as expected from the branching ratio of photomeson process. In the case of $\gamma_{p,\max}'=10^5$, the BH process may dominate the high-energy prompt emission, and hence the neutrino flux that is related to the photomeson process is lower than the high-energy photon flux. Also, due to a lower maximum proton energy, the neutrino flux peaks at a lower energy. On the other hand, the overall interaction efficiency in the case of $\gamma_{p,\max}'=10^5$ is smaller than the case of $\gamma_{p,\max}'=10^6$, so the required baryonic loading factor in the former case is larger than in the latter case. Thus, the neutrino fluxes in the two cases, as is shown in Fig.~\ref{fig:neutrino}, result in a similar neutrino detection probability by IceCube. Given the effective area of IceCube and the zenith angle of the GRB, and assuming a flavor ratio of 1:1:1 after oscillation, we find the probability of detecting a muon neutrino or anti-muon neutrino in the range of 10\,TeV$-$10\,PeV is 0.001 for GRB~090926A, 0.03 for GRB~090902B in both cases of $\gamma_{p,\max}'$, and 0.01 in the case of $\gamma_{p,\max}'=10^5$ and 0.008 in the case of $\gamma_{p,\max}'=10^6$ for GRB~080916C. Provided that ten GRBs of similar properties to these three kinds of GRBs are detected by LAT per year, we expect $<1$ GRB-correlated neutrinos can be detected for past 9 years (2008-2017), which is consistent with the observation. But the persistent accumulative neutrinos flux from the future LAT GRBs would lead to the detection of correlation between neutrino event and LAT GRBs in another $\sim 10$ years. The detection or non-detection of neutrinos for LAT GRBs in the future could be a test for the hadronic model.
However, for the most majority of GRBs, we expect a quite different set of parameters especially a much smaller baryon loading factor, which should basically follow the constraint from the non-detection of GRB-neutrino correlation by IceCube \citep{aartsen15, aartsen16, aartsen17}. This was also found from early theoretical studies \citep{li12,he12,li13,zhang13,Liu13b}.

%

We also note that the constraints on GRB jet parameters from IceCube is based on the assumption that $\mathcal{A}\sim 1$. For most GRBs in which high-energy emission is not detected. However, as we have shown in this paper, we do not need to invoke a large maximum proton energy to explain the GeV emission with the hadronic model. Thus, the maximum proton energy or Lorentz factor could be as low as $10^{4}$. In Fig.~\ref{fig:neutbench}, we present the neutrinos flux from a typical GRB of $L_\gamma=10^{51}$erg/s under different maximum proton energy, with all the other parameters the same with the benchmark parameters. Because the photomeson efficiency drastically decreases with the proton energy, we can see both the peak energy and the peak flux of neutrino decrease with decreasing $\gamma_{p,\max}'$. Comparing the neutrino flux in the case of $\gamma_{p,\max}'=10^7$ (corresponding to $\mathcal{A} \simeq 220$) to the case of $\gamma_{p,\max}'=10^4$ (corresponding to $\mathcal{A} \simeq 9.2\times 10^5$), the peak flux is two orders of magnitude lower than the former one. And the predicted neutrino event in the latter case is less than that in the former case by more than one order of magnitude. Thus, with a low maximum proton energy, the current constraints on GRB parameters such as the baryonic loading factor and the bulk Lorentz factor could be relaxed to certain extent.


\subsection{low-energy excess due to cascade emission}
The synchrotron radiation of secondary $e^\pm$ may lead to a low-energy excess in the spectrum if the low-energy spectrum of the Band component is hard (i.e., a large $\alpha$), such as in GRB~090902B. This feature could be an indication of the baryon component in the composition of a GRB jet \citep{asano10}. However, for most of GRBs, the low-energy Band spectrum is not hard enough so that the Band component would conceal such a component above the threshold energy of the current detectors (e.g., \emph{Fermi}-GBM at 8keV) even it really exists. Future GRB detectors such as the SVOM satellite \citep{Cordier15} have a lower threshold energy of 4\,keV and thus provides a better chance to reveal such a low-energy excess. If the synchrotron radiation extends to even lower energy such as the optical band, there may be a chance to capture it by GWAC which is the ground-based system of the SVOM mission. But we should be cautious because the synchrotron radiation of the secondary $e^\pm$ may undergo self-absorption below certain energy and the optical emission of GRBs may stem from some other mechanisms.

\section{Summary}
To summarize, we revisited the hadronic model for the high-energy emission of GRBs in the prompt emission phase. While the energy of accelerated protons in GRB jets is assumed to reach $\sim 10^{20}$eV in the previous literatures, we show that, in the case of a lower maximum proton energy, the photomeson process may be less important than the Bethe-Heitler process in converting energies of protons to gamma-rays via an electromagnetic cascade. Taking GRB~090926A, GRB~090902B and GRB~080916C as examples, we found that the synchrotron radiation and inverse Compton scattering of secondary $e^\pm$ from the proton-induced cascade via the Bethe-Heitler process and photomeson process can reproduce various types of the observed spectrum of the prompt high-energy emission of GRBs with a relatively low maximum proton energy. The adopted parameters in the spectrum fittings are consistent with the constraints from the null detection of GRB-correlated neutrino events by the IceCube neutrino telescope. The cascade emission may also lead to a low-energy excess below a few keV and might be used as an indication for the baryon component in the GRB ejecta.

\acknowledgements
We thank Prof. Aharonian, Felix for his invaluable discussion, and the anonymous referee for the useful comments. This work is supported by the National Basic Research Program (``973'' Program) of China (grant No. 2014CB845800), the National Key Research and Development
Program of China (grant No. 2017YFA0402600), and the National Natural Science Foundation of China (grant No. 11573014). This work is also partly supported by the joint research program of the Institute for Cosmic Ray Research (ICRR), The University of Tokyo, and Grants-in-Aid for Scientific Research nos. 15K05069, 16K05291 (K.A.) from the Ministry of Education, Culture, Sports, Science and Technology (MEXT) of Japan.

\clearpage

\appendix

\section{Solving the spectrum of cascaded electrons in quasi-steady state}
Basically, we follow the treatment in \citet{bottcher13} for blazars, but also consider the additional BH process for protons and IC radiation for electrons which are important in the scenario of GRB jets. The key to obtain the cascaded electron spectrum in the quasi-steady state is to solve the following equation
\begin{equation}
{{n'}_e}({{\gamma '}_e}) = -\frac{1}{{{\dot{\gamma}'}_e}}\int_{{{\gamma '}_e}}^\infty  d {{\tilde \gamma }_e'}[{{Q_e}({{\tilde \gamma}_e'}) + {\dot n'}_{e,\gamma \gamma } ({{\tilde \gamma }_e'})}],
\label{equne}
\end{equation}
The cooling rate of electrons via synchrotron radiation and IC scattering can be given by
\begin{equation}
{{\dot \gamma '}_e} =  - \frac{{c{\sigma _T}{{B'}^2}}}{{6\pi {m_e}{c^2}}}{{\gamma '}_e}^2 + {{\dot \gamma' }_{e,\rm IC}},
\end{equation}
where ${{\dot \gamma' }_{e,\rm IC}}$ accounts for the IC cooling including the Klein-Nishina effect, which can be calculated following \cite{blumenthal70}.  ${{Q_e}({{ \gamma '}_e})}$ is the injection rate of the first-generation electrons (including both $e^-$ and $e^+$) from the photomeson process and the BH process, which are calculated based on the semi-analytical treatment in \cite{kelner08}. ${{{\dot n'}_{e,\gamma \gamma }}({{\tilde \gamma '}_e})}$ is the pair production rate through the $\gamma\gamma$ annihilation, including the annihilation of the high-energy photon from the neutral pion decay produced in the photomeson process, and the high-energy photon produced by the synchrotron radiation and the IC scattering, i.e.,
\begin{equation}
{{\dot n'}_{e,\gamma \gamma }}({{\gamma '}_e}) = {f_{abs}}({\varepsilon _1})(\dot n_{{\varepsilon _1}}^0 + \dot n_{{\varepsilon _1}}^{sy} + \dot n_{{\varepsilon _1}}^{\rm IC}) + {f_{abs}}({\varepsilon _2})(\dot n_{{\varepsilon _2}}^0 + \dot n_{{\varepsilon _2}}^{sy} + \dot n_{{\varepsilon _2}}^{\rm IC}),
\label{ggab}
\end{equation}
with
\begin{equation}
{f_{abs}}(\varepsilon ) = 1 - \frac{{1 - {e^{ - {\tau _{\gamma \gamma }}(\varepsilon )}}}}{{^{{\tau _{\gamma \gamma }}(\varepsilon )}}}
\end{equation}
being the absorbed fraction of photons. Eq.~(\ref{ggab}) contains two parts, since the $\gamma\gamma$ annihilation produce two electrons, taking a fraction $f_\gamma$ and $1-f_\gamma$ of the energy of initial high energy photon, respectively. If the center-of-mass energy of the interaction is sufficiently high, one of the outgoing electron will take most of the energy of the initial photon. According to \citet{bottcher13}, $f_\gamma=0.9$ is a good agreement with the numerical Monte Carlo simulations. As a result, to the produce an electron with energy ${{\gamma '}_e}$, the photons should have the energy of either ${\varepsilon _1} = {{\gamma '}_e}/{f_\gamma}$, or ${\varepsilon _2} = {{\gamma '}_e}/(1 - {f_\gamma })$. In Eq.~(\ref{ggab}), $\dot n_{{\varepsilon }}^0$ can be also found by the analytical treatment in \citet{kelner08}. $\dot n_{{\varepsilon}}^{sy}$ can be described as
\begin{equation}
\dot n_\varepsilon ^{sy}{\text{ = }}{A_0}{\varepsilon ^{ - 2/3}}\int_1^\infty  {d{{\gamma '}_e}} {{n'}_e}({{\gamma '}_e}){\gamma '}_e^{ - 2/3}{e^{ - \varepsilon /(b {\gamma '}_e^2)}},
\end{equation}
with
\begin{equation}
{A_0} = \frac{{c{\sigma _T}{{B'}^2}}}{{6\pi {m_e}{c^2}}}\frac{1}{{\Gamma (4/3){b^{4/3}}}},
\end{equation}
where $\Gamma (4/3) = {\text{0}}{\text{.89297}}$, $b = B'/{B_{\rm crit}}$ and ${B_{\rm crit}} = 4.4 \times {10^{13}}\,\rm G$, while $\dot n_{{\varepsilon }}^{\rm IC}$ can be given by
\begin{equation}
\dot n_\varepsilon ^{\rm IC} = \int_1^\infty  {d{{\gamma '}_e}} {{n'}_e}({{\gamma '}_e})\frac{1}{{{{\gamma '}_e}{m_e}{c^2}}}\frac{{dN}}{{dtd{E_1}}},
\end{equation}
where $\frac{{dN}}{{dtd{E_1}}}$ is given by equation (2.48) of \cite{blumenthal70}.

Since the electron spectrum ${{n'}_e}({{\gamma '}_e})$ exists on both sides of equation (\ref{equne}), this equation can be evaluated progressively, starting from the highest electron energies and then using the solution of ${{n'}_e}({{\gamma '}_e})$ for large $\gamma'_e$ as one progresses toward the lower values of $\gamma'_e$, to obtain the equilibrium pair distribution ${{n'}_e}({{\gamma '}_e})$, which has an excellent agreement with the results of the Monte-Carlo simulations \citep{bottcher13}. Then, one can obtain the synchrotron and IC spectra from the equilibrium pair distribution after the absorption by the target photon field. Since the characteristic energy for the absorption of GeV photons by photon-photon annihilation is around tens of MeV, where the contribution of the extra component could be higher than that of the Band component, to get the final and correct spectra of photons, we execute an iteration procedure until the self-consistent results after the photon-photon absorption are reached.

\clearpage

\begin{figure}
\epsscale{.96} \plotone{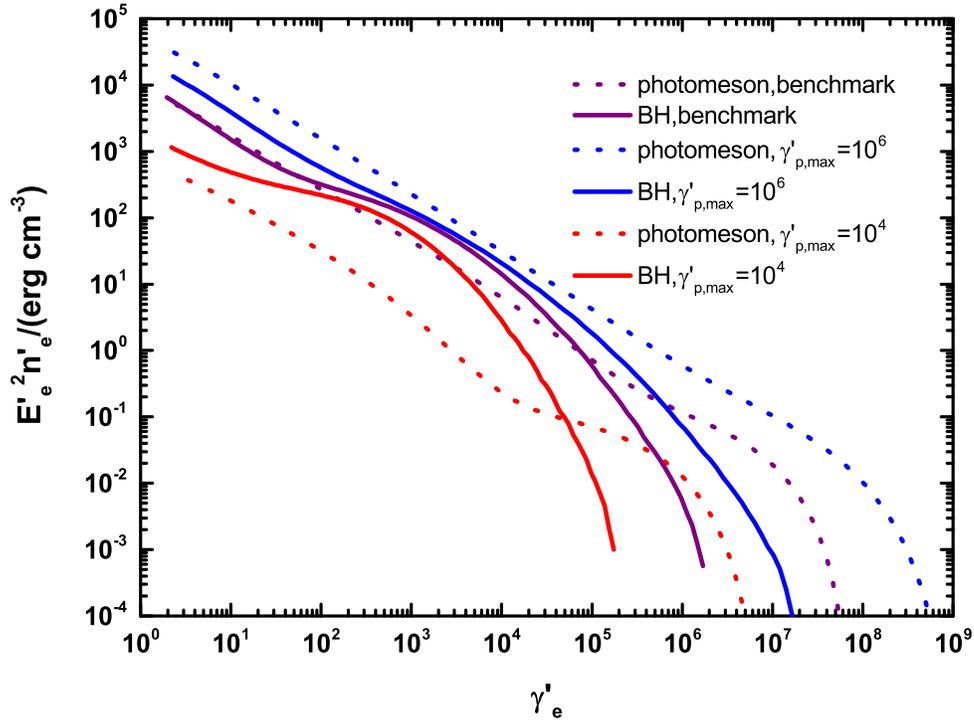}\caption{Quasi-steady-state spectrum of the cascaded electrons in the comoving frame of the GRB jet with fiducial parameters (except for the maximum proton Lorentz factor). Solid curves show the electrons originating from the BH process while dashed curves show the electrons originating from the photomeson process. Red, purple, blue colors are for $\gamma_{p,\max}'=10^4,10^5$ and $10^6$ respectively.}\label{fig:espec}
\end{figure}

\begin{figure}
\epsscale{.96} \plotone{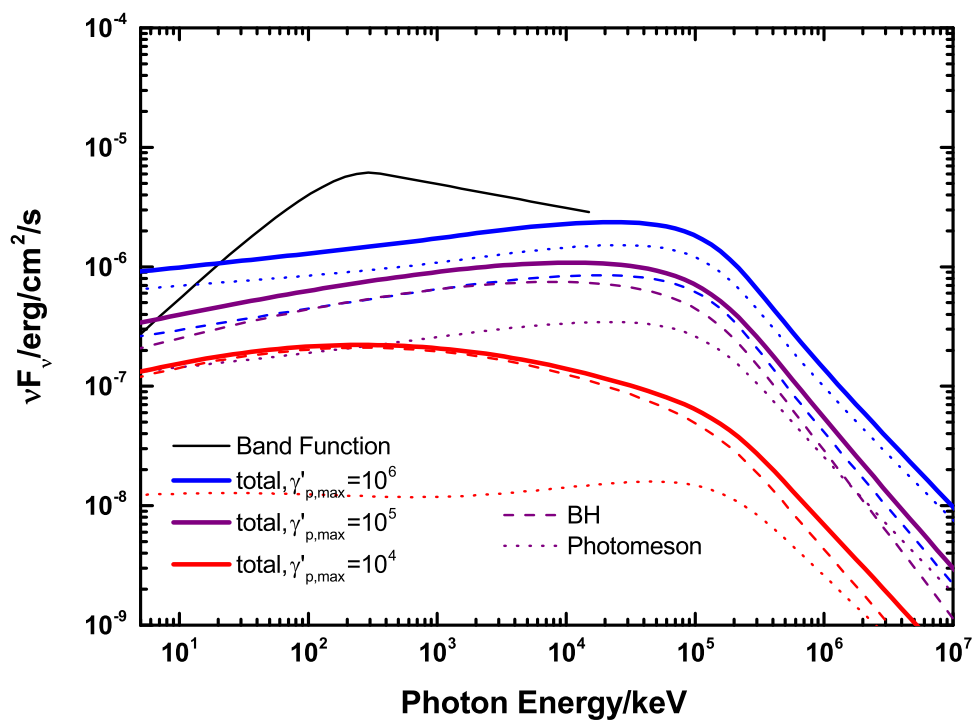} \caption{Spectrum of the cascade emissions in a GRB jet with the benchmark parameters (except the maximum proton Lorentz factor $\gamma_{p,\max}'$).  The dashed and dotted curves are corresponding to the Bethe-Heitler and photomeson processes respectively. \label{fig:gmax}}
\end{figure}

\begin{figure}
\epsscale{.96} \plotone{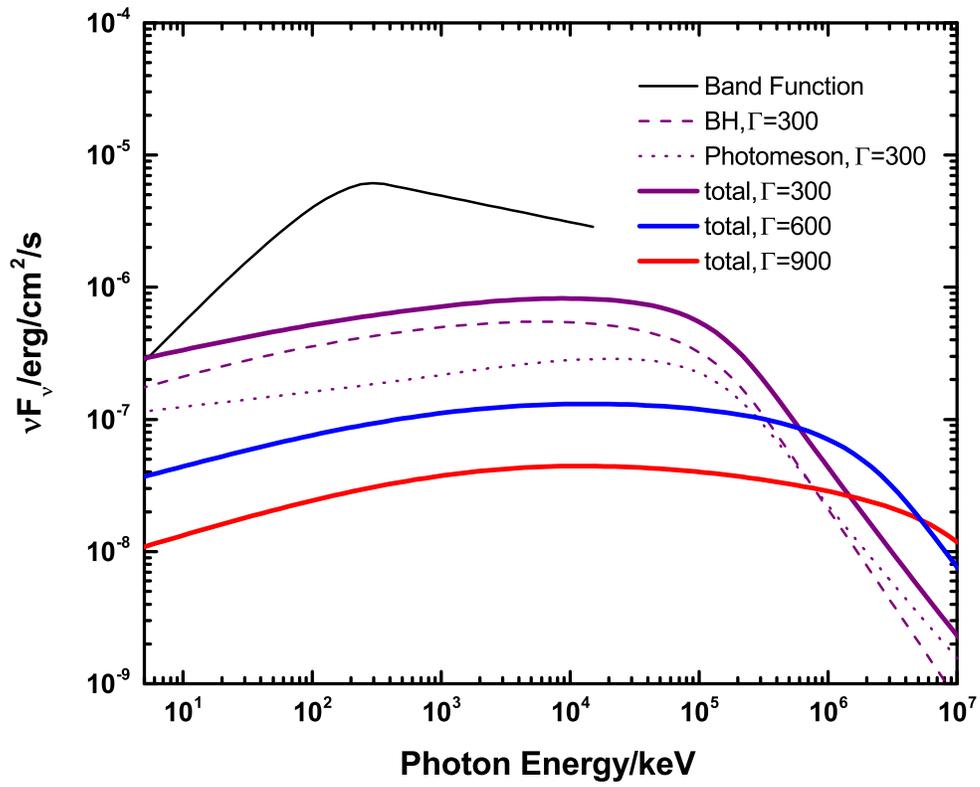} \caption{The same as Fig.~\ref{fig:gmax} but for different bulk Lorentz factor of GRB jet $\Gamma$. \label{fig:LF}}
\end{figure}

\begin{figure}
\epsscale{.96} \plotone{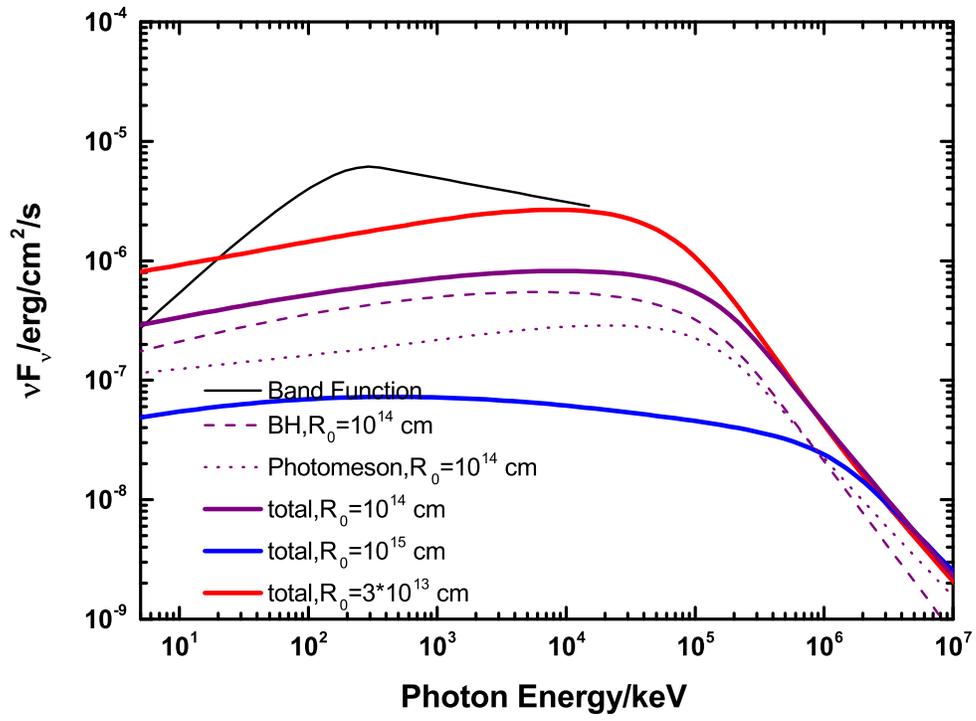} \caption{The same as Fig.~\ref{fig:gmax} but for different dissipation radius of GRB jet $R$.\label{fig:radius}}
\end{figure}

\begin{figure}
\epsscale{.96} \plotone{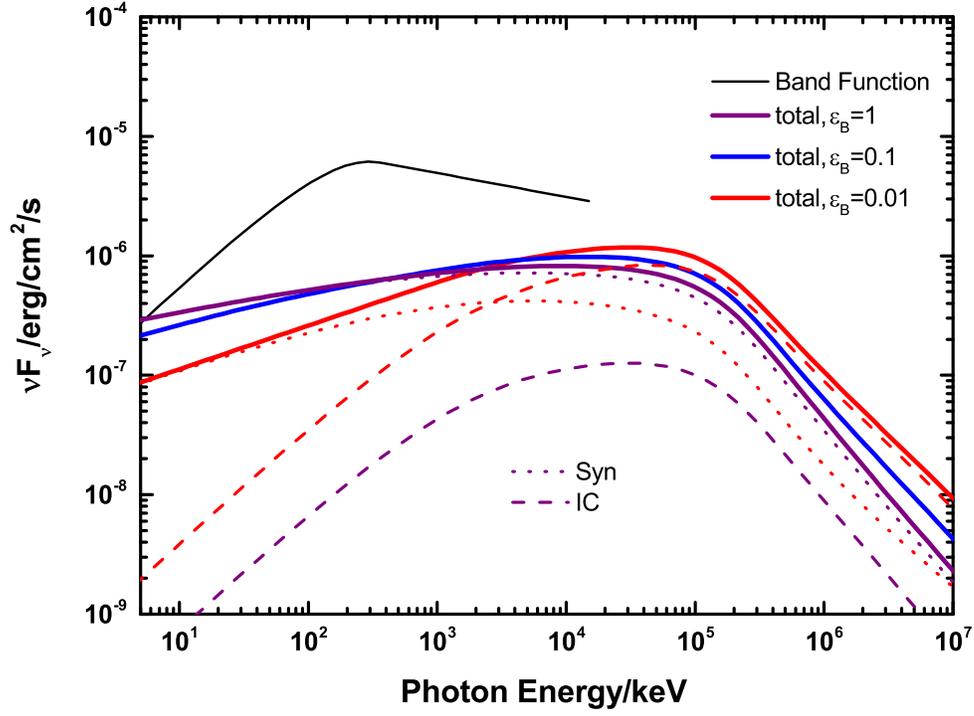} \caption{The same as Fig.~\ref{fig:gmax} but for the equipartition coefficient for magnetic field in GRB jet $\epsilon_B$, while the dashed and dotted lines are corresponding to the IC radiation and synchrotron radiation respectively.\label{fig:epB}}
\end{figure}

\begin{figure}
\epsscale{.96} \plotone{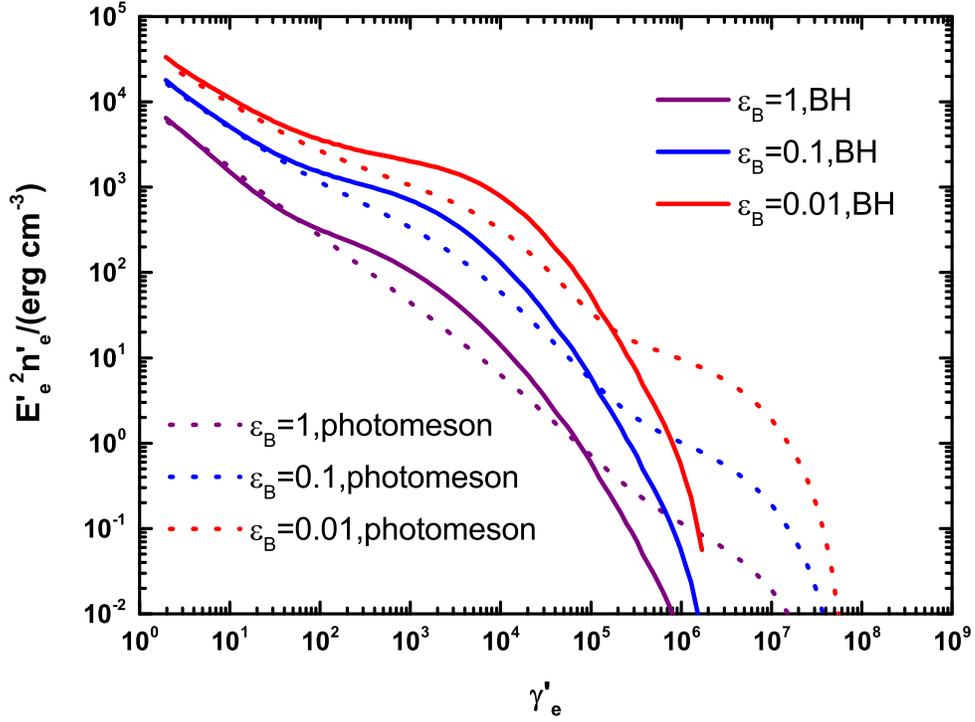} \caption{Quasi-steady-state spectrum of the cascaded electrons in the comoving frame of the GRB jet with fiducial parameters (except for the equipartition coefficient of magnetic field). Purple, blue and red curves are for $\epsilon_B=1,0.1$ and $0.01$ respectively, while solid and dashed curves are the contribution by the BH process and the photomeson process respectively.\label{fig:espec_epB}}
\end{figure}

\begin{figure}
\includegraphics[width=0.7\textwidth]{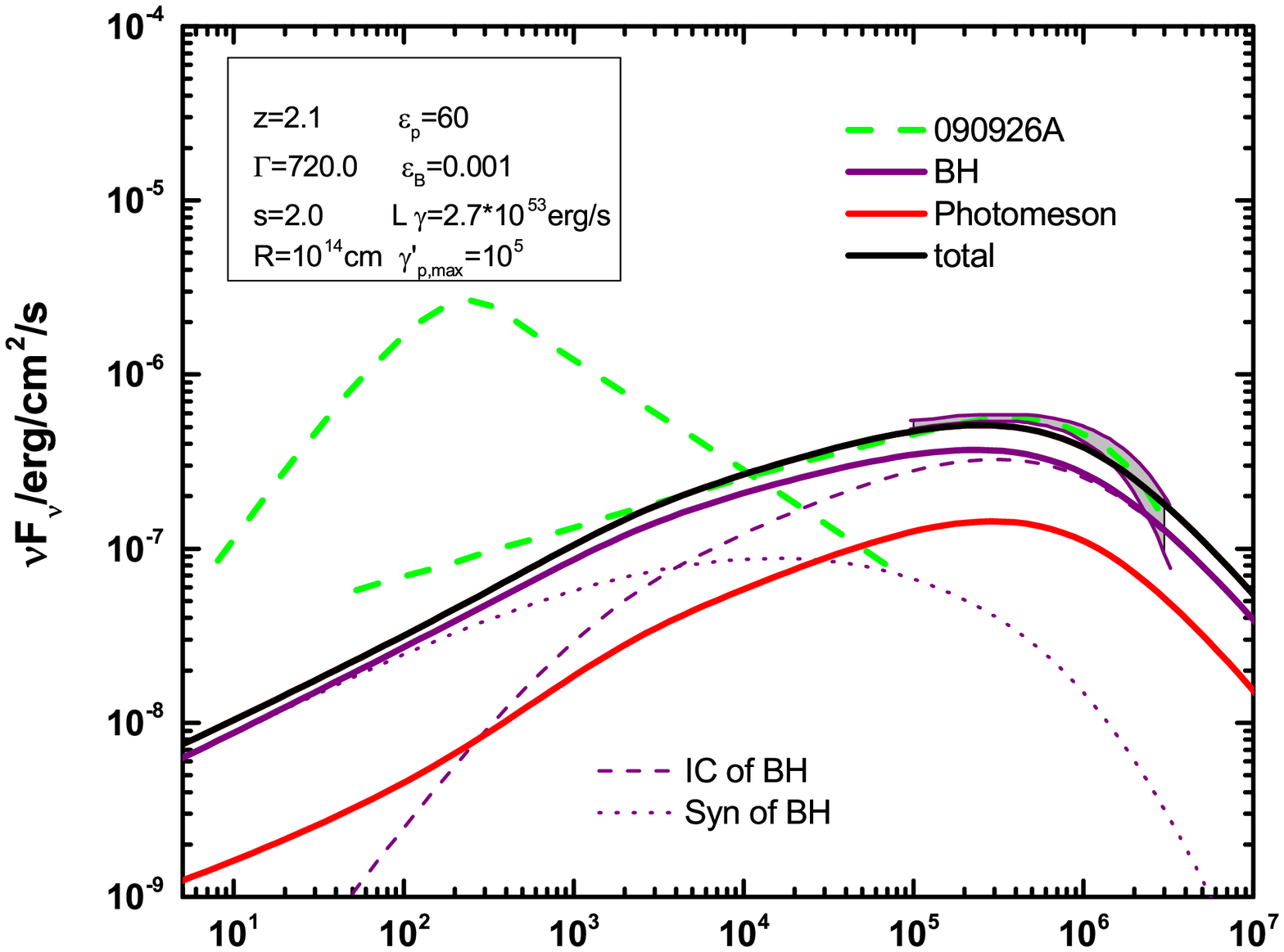}
\includegraphics[width=0.7\textwidth]{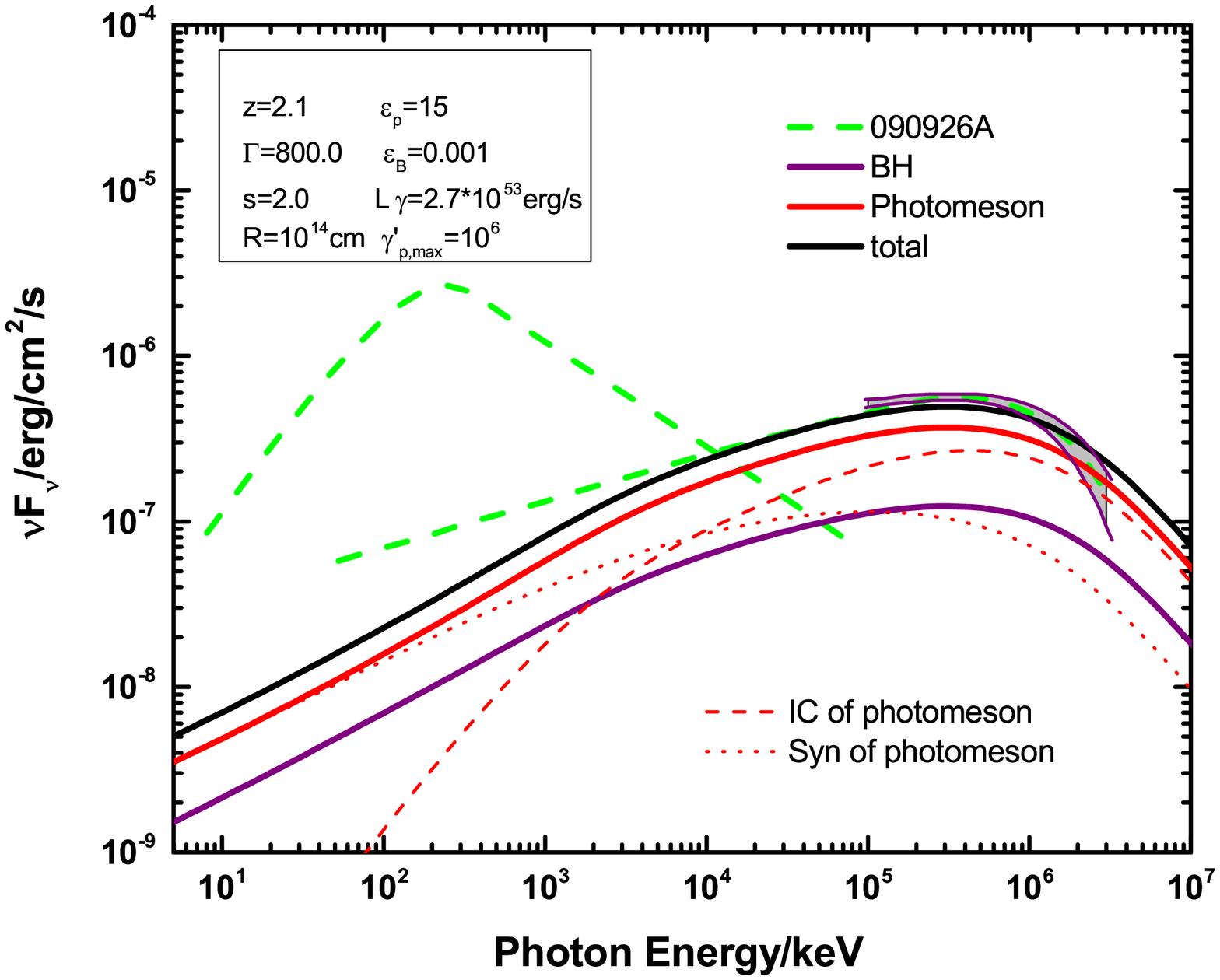}
\caption{The spectral fitting for GRB 090926A with $\gamma'_{p,\max}=10^5$ ({\bf top}) and $\gamma_{p,\max}'=10^6$ ({\bf bottom}). The purple curves represent the emissions of cascaded electrons originating from BH process while the red ones represent that from the photomeson process. The red or purple dotted curves and dashed curves show the synchrotron radiation and IC radiaiton of the electrons from the dominant process. The green dashed curves show the fitting of the burst's spectrum by Fermi-LAT collaboration. The shaded region shows the uncertainty in the spectrum fitting at high energy end.\label{fig:090926A}}
\end{figure}

\begin{figure}
\includegraphics[width=0.7\textwidth]{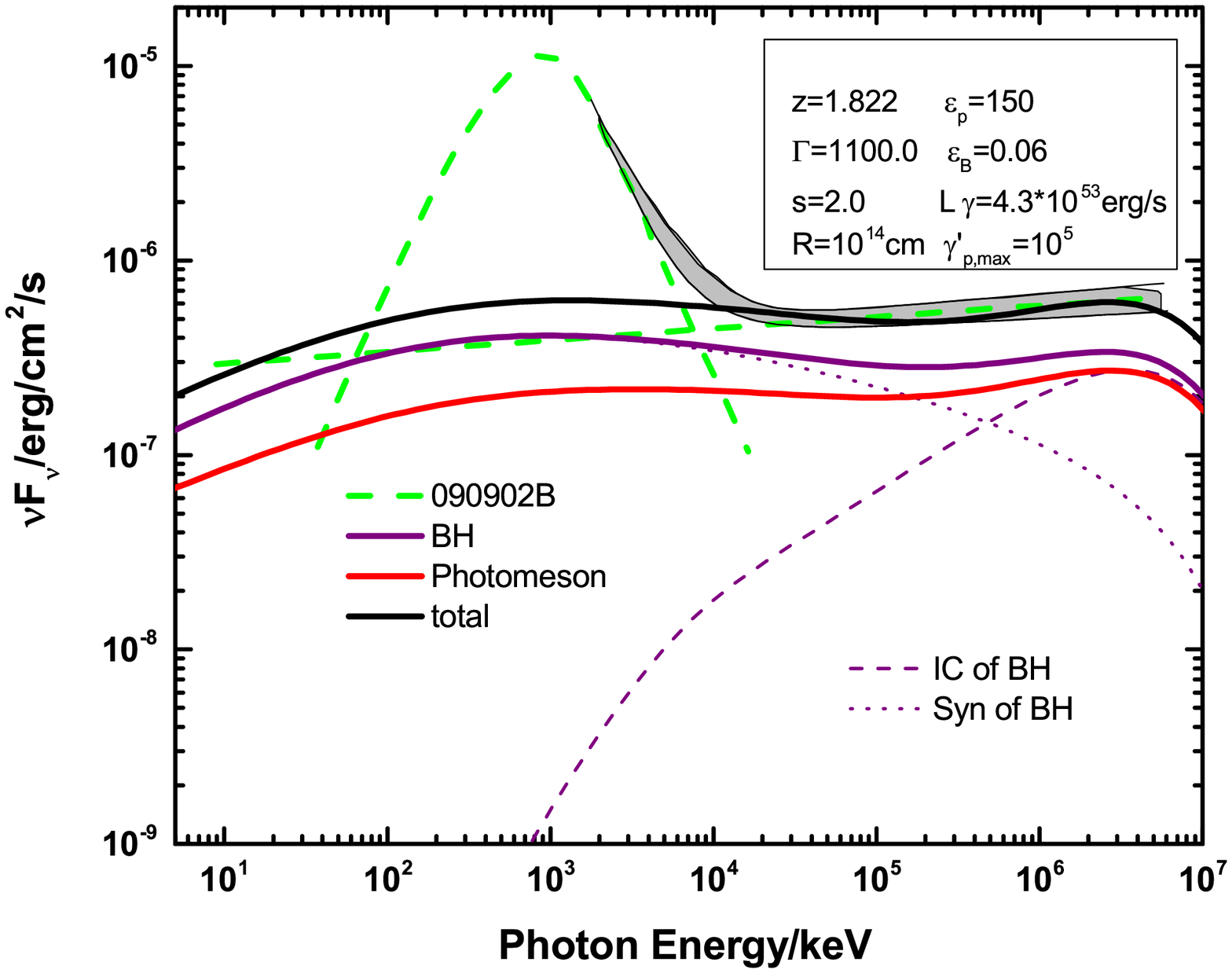}
\includegraphics[width=0.7\textwidth]{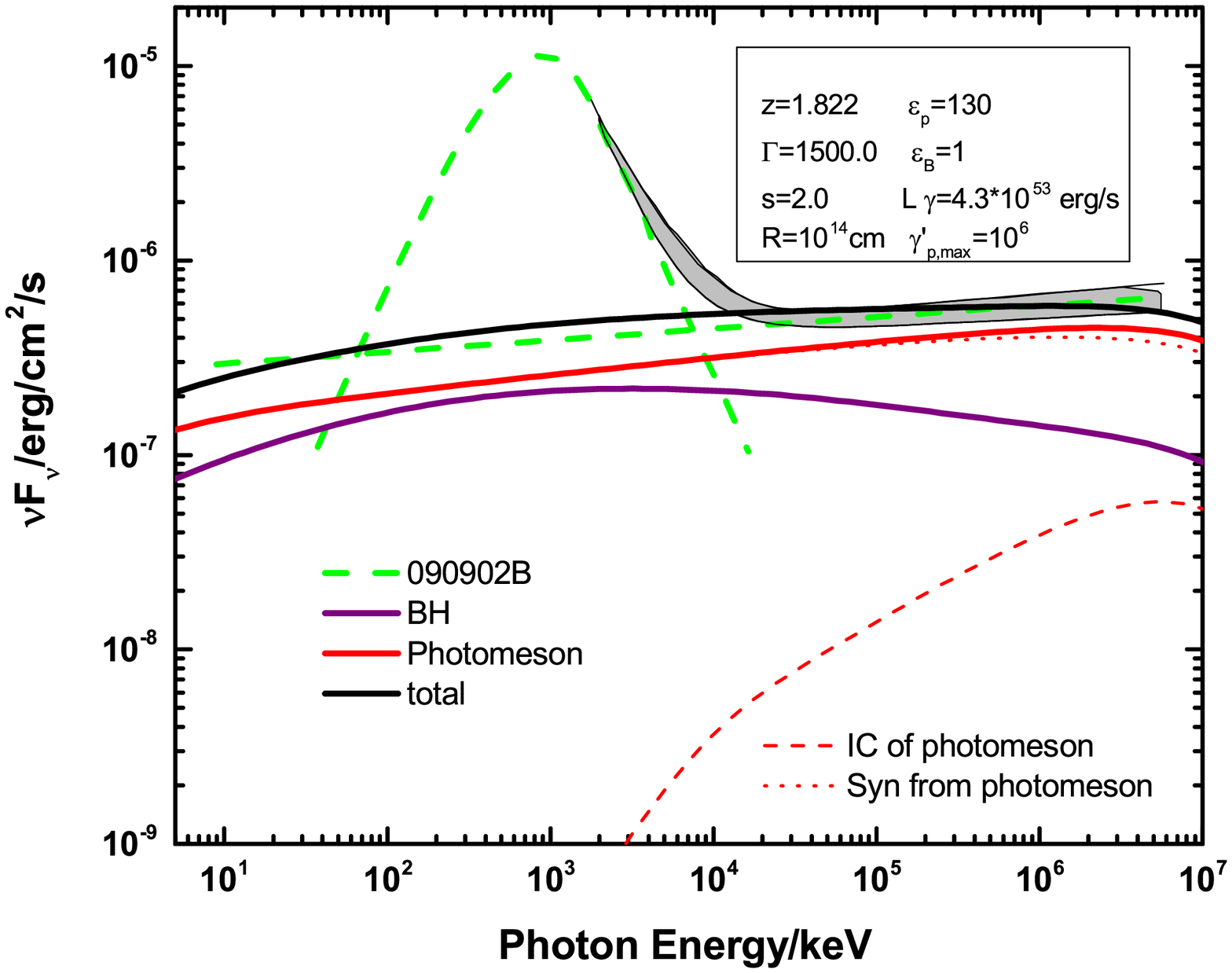}
\caption{The same as Fig.~\ref{fig:090926A} but for GRB~090902B.\label{fig:090902B}}
\end{figure}

\begin{figure}
\includegraphics[width=0.7\textwidth]{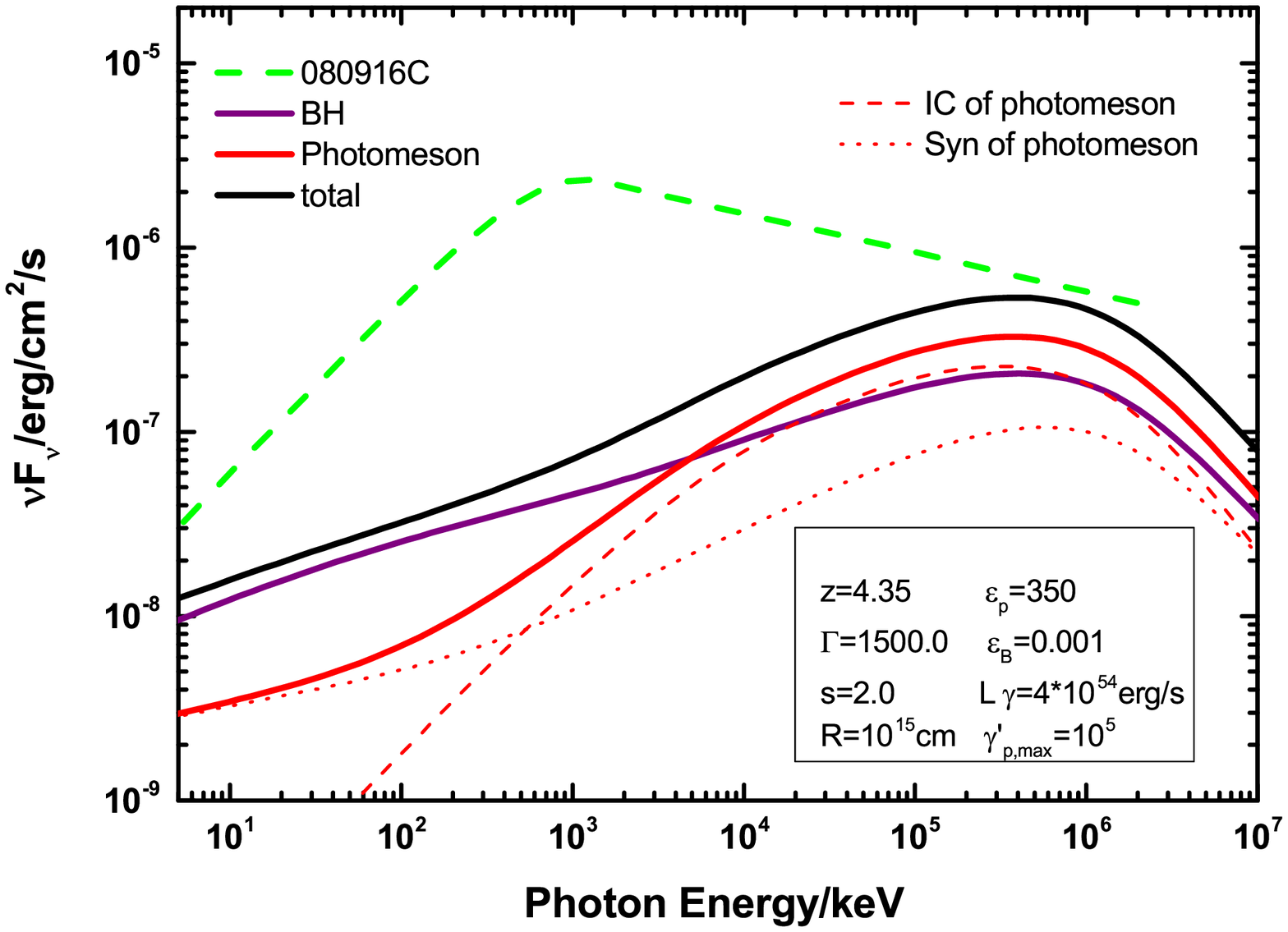}
\includegraphics[width=0.7\textwidth]{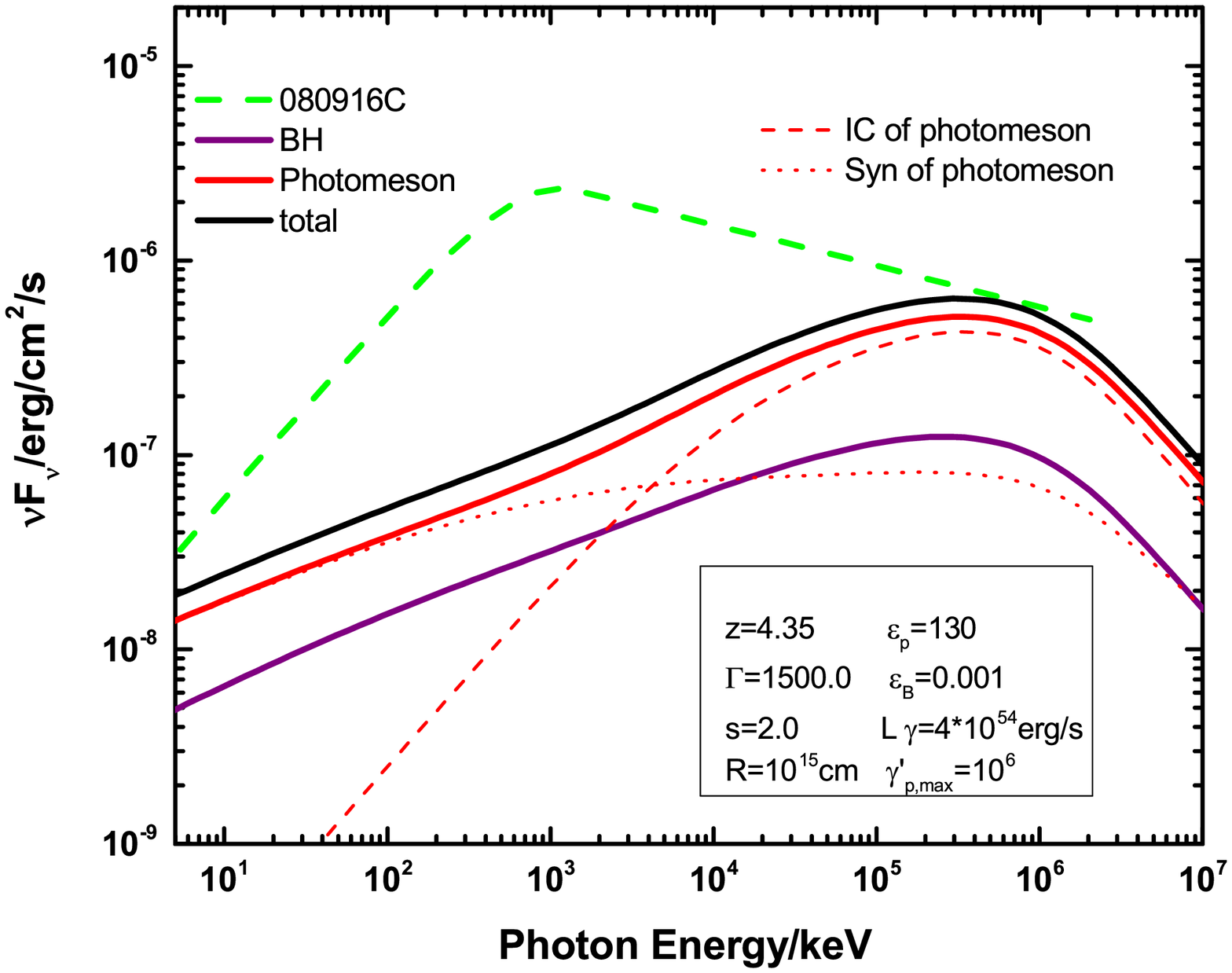}
\caption{The same as Fig.~\ref{fig:080916C} but for GRB~080916C.\label{fig:080916C}}
\end{figure}

\begin{figure}
\includegraphics[width=1\textwidth]{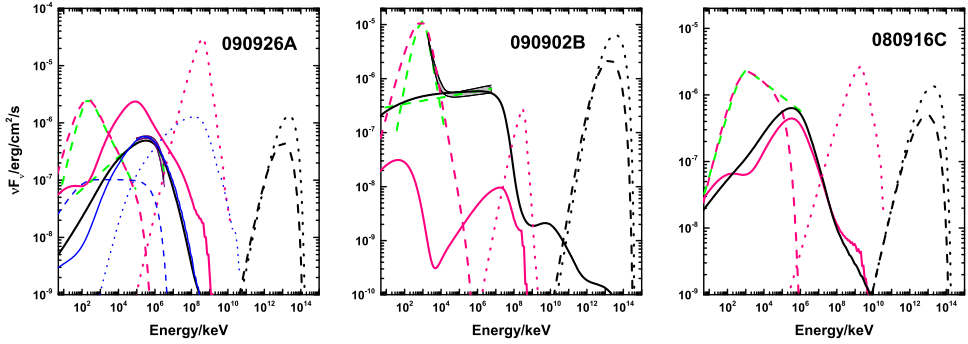}
\caption{The cascade emissions initiated by SSC emissions of primary electrons for the GRB 090902B, 090926A and 080916C (pink solid curves) by normalizing the corresponding synchrotron emission (pink dashed curves) with the Band component (see discussion in Section 4.1). The unabsorbed SSC emissions are shown as pink dotted curves. The injection rate of the first-generation secondary photons and electrons from photomeson process are shown as black dotted and black dashed curves respectively, while their initiated cascade emissions in quasi-steady state are shown as black solid curves. Note that the component shown with pink dotted curves, black dotted curves and black dashed curves can not be observed since they will have interactions and initiated EM cascade inside the GRB jet. They are presented here just to show the injection rate for the respecitve cascades. In all three panels, the parameters of GRBs are exactly same with those used in the lower panels in Fig.~\ref{fig:090926A}-\ref{fig:080916C}. Some additional paramters for the primary eletron distributions, i.e., the break Lorentz factors in the electron spectrum, the low-energy spectral slope and the high-energy spectral slope, are set to $4883$, $0.82$ and $-4.3$ for GRB 090926A,  $1643$, $2.14$ and $-6.80$ for GRB 090902B, and $2.21\times 10^4$, $0.04$ and $-3.42$ for GRB 080916C, respectively. The blue solid curve in the leftmost panel represents the cascade emission from a co-accelerated electron component, with $\epsilon_e=1$. See text for details. \label{fig:ssc}}
\end{figure}

\begin{figure}
\epsscale{.96} \plotone{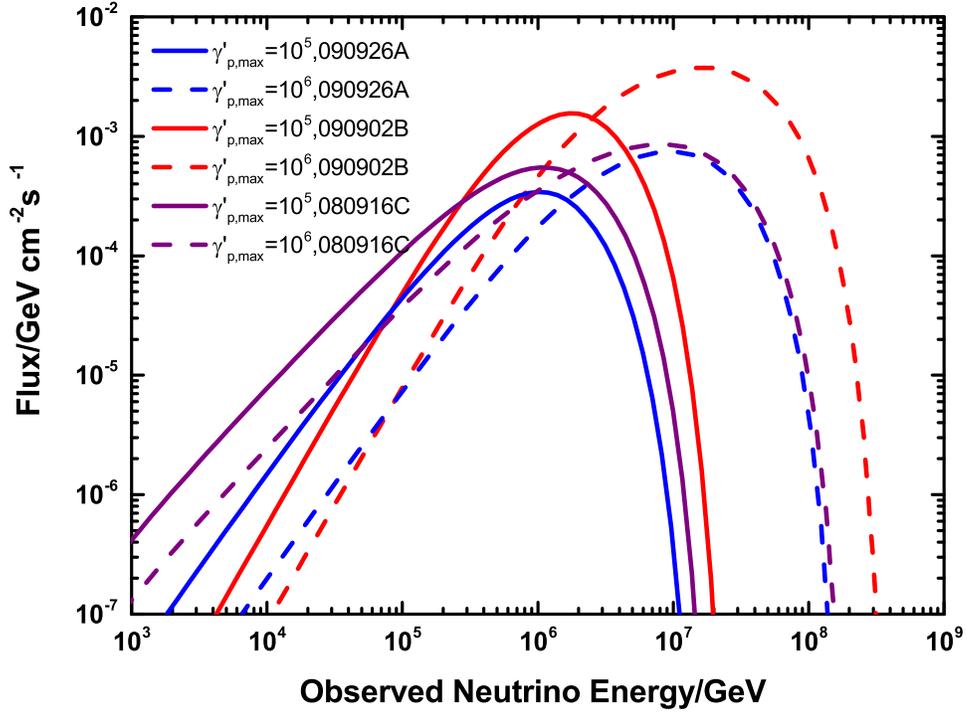} \caption{All-flavor neutrino flux for GRB~090926A (blue), GRB~090902B (red) and GRB~080916C (purple) obtained under the same parameters employed in Fig.~\ref{fig:090926A}, Fig.~\ref{fig:090902B} and Fig.~\ref{fig:080916C} respectively.\label{fig:neutrino}}
\end{figure}

\begin{figure}
\epsscale{.96} \plotone{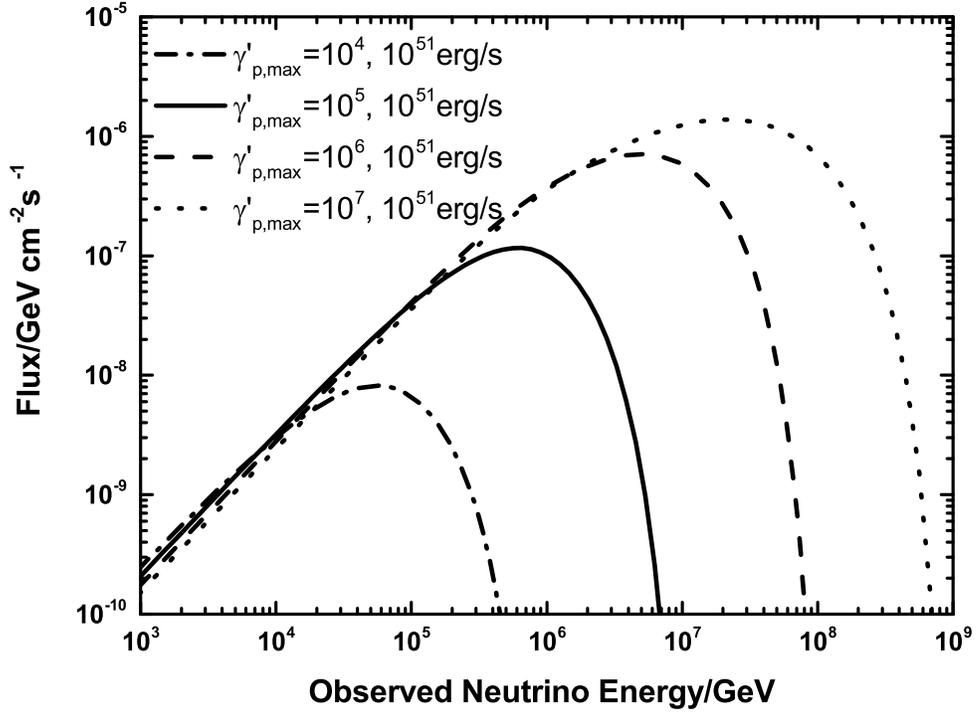} \caption{All-flavor neutrino flux from a GRB of the typical luminosity $L_\gamma=10^{51}\,\rm erg/s$ (in 1\,keV--10\,MeV) with different maximum proton energy (Lorentz factor) in the comvoing. Other parameters are the same with the benchmark parameters listed in Table.~1.\label{fig:neutbench}}
\end{figure}

\end{document}